\documentclass[12pt]{article}
\newcommand{\blind}{0}

\usepackage{times,color,verbatim,graphicx,amssymb,lscape,amsmath}
\usepackage{graphicx,graphics,psfrag,epsf}
\usepackage{enumerate}
\usepackage{url} % not crucial - just used below for the URL
\usepackage{multirow}
\usepackage{multicol}
\usepackage{rotating}
\usepackage{relsize}
\usepackage{bm}
\usepackage{hyperref}
\usepackage{multicol}
\usepackage{setspace}
\usepackage{enumerate}
\usepackage{color}
\usepackage[authoryear,sort]{natbib}
\usepackage{footnote}
\usepackage{mymacros_ishd}

\date{}
\begin{document}
\def\spacingset#1{\renewcommand{\baselinestretch}%
{#1}\small\normalsize} \spacingset{1}
\if0\blind
{
  \title{\bf Post Selection Shrinkage Estimation for High Dimensional Data Analysis}
  \author{Xiaoli Gao\\
    Department of Mathematics and Statistics\\
     University of North Carolina at Greensboro\\
    and \\
    S. E. Ahmed  \\
   Department of Mathematics\\
    Brock University\\
    and\\
    Yang Feng\\
   Department of Statistics\\
    Columbia University}
  \maketitle
} \fi

\if1\blind
{
  \bigskip
  \bigskip
  \bigskip
  \begin{center}
    {\LARGE\bf Post Selection Shrinkage Estimation for High Dimensional Data Analysis}
\end{center}
  \medskip
} \fi

\bigskip

\begin{abstract}
%%%%%%%%%%%%%%%%%%%%%%%%%%%%%%%%%%%%%%%%%%%%%%%%
%%%%%%%%%%%%%%%%%%%%%%%%%%%%%%%%%%%%%%%%%%%%%%%%
%\section*{Abstract}
  In   high-dimensional data settings where $p\gg n$,
  many penalized regularization approaches were
studied for simultaneous variable selection and estimation.
However, with the existence of covariates with weak effect,
many existing variable selection methods, including Lasso and its generations,
 cannot  distinguish covariates with
weak and no contribution.
 Thus, prediction based on a subset model of  selected covariates only
 can be inefficient.
 In this paper, we propose a post selection shrinkage
 estimation strategy to improve the prediction performance
of a selected subset model.
 Such a post selection shrinkage  estimator (PSE) is data-adaptive and
  constructed by shrinking  a post selection weighted ridge  estimator
 in the direction of a selected candidate subset.
%A HD pretest ridge estimator was also constructed as a by-product3.
 Under an asymptotic distributional quadratic risk criterion, its prediction
 performance is explored analytically. We show that the
proposed post selection PSE performs better than
the post selection weighted ridge estimator.
More importantly,
%when a candidate subset model from
%any Lasso-type method is under-fitted,
it improves the prediction performance
of any candidate subset model selected from
most existing Lasso-type variable selection methods significantly.
The relative performance of the post selection PSE
is demonstrated by both simulation studies and
 real data analysis.
\end{abstract}

\noindent%
{\it Keywords:}   Asymptotic risk; Lasso; Ridge regression; (Positive) shrinkage estimation;
Post selection; Sparse model.

\vfill

\newpage
\spacingset{1.45} % DON'T change the spacing!

\section{Introduction}

Many high-dimensional data  arise in  biological, medical, social, and economical
studies.  Due to the  trade-off between  model complexity and model prediction,
the statistical inference of model selection
 becomes extremely important and challenging in high-dimensional  data analysis.
Consider a classical high-dimensional  linear regression model with  observed
response variable $y_i$ and covariates $x_i$s, %$ \bx_i=(x_{i1},\cdots,x_{ip_n})'$ as follows,
\bel{linear model}
y_i=\sum_{j=1}^{p_n} x_{ij} \beta_j+\veps_i,\quad 1\le i\le n,
\eel
where % $\bbeta_n=(\beta_1,\cdots,\beta_{p_n})'$ is a $p_n$-dimensional
%vector of the unknown parameters, and
$\veps_i$s are  independent and identically distributed random errors
with center $0$ and variance $\sigma^2$.
Without loss of generality, we do not include the intercept
in the model by assuming all data have been centered.
Here the subscript $n$ in $p_{n}$ indicates that the number of coefficients
may increase with the sample size $n$. Such a notation will be
used throughout the paper without further explanation.
%In model \eqref{linear model},  the number of covariates, $p$ may increase with the sample size $n$.

 Over the past two decades, many penalized regularization approaches have been
 developed  to do variable selection and estimation simultaneously.
 Among them,  the Lasso  \citep{tibshirani:1996} is one of the most popular
 approaches due to its convexity and computation
 efficiency.
%The Lasso is designed to minimize a quadratic loss function subject to a bound on
%the $\ell_1$ norm of the coefficients.
%Due to the non-differentiability of the $\ell_1$ norm,
%Lasso can produce a sparse solution, meaning  exact zero estimation for some coefficients.
In general, the Lasso penalty turns to select an
over-fitted model since it penalizes
 all coefficients equally
\citep{leng.lin.ea:2006}.
  Many endeavors have been undertaken to improve the Lasso to
reach both the variable selection consistency and the estimation consistency.
To list a few, the Adaptive Lasso \citep{zou:2006}, SCAD \citep{fan.li:2001}, and MCP \citep{zhang:2010}, among others.

In order to have nice estimation and selection properties,
most Lasso-type penalties make some important assumptions on both
the true model and the designed covariates.
For example,  the true model  is often assumed to be
sparse such that (i) most $\beta_j$s are zeros except for a few ones and
(ii)  all those nonzero $\beta_j$'s are larger
than an inflated noise level,  $c\sigma\sqrt{(2/n)\log(p_n)}$ with $c\ge 1/2$
\citep{zhang.Zhang:2014}.
Additional  assumptions made on the designed covariates include
the adaptive irrepresentable condition and the restricted eigenvalue conditions.
For detailed information, we refer to \cite{zhao.yu:2006}, \cite{huang.ma.zhang:2008},
and \cite{bickel.ritov.ea:2009}.

However, those conditions are somewhat restrictive and
are not judiciously justified in real applications.
Consequently,  Lasso and its generalizations may have lower prediction efficiency
once those assumptions are violated.
To fix the idea, we take the sparse model assumption (ii) as an example.
Suppose we can divide the index set $\{1,\cdots, p_n\}$ into
three disjoint subsets:  $S_{1}$, $S_{2}$ and $S_{3}$. In particular,
$S_{1}$ includes indexes of nonzero
$\beta_i$'s which are moderately large and  easily  detected;
 $S_{3}$ includes indexes with only zero coefficients;
 $S_{2}$, being the intermediate,
  includes indexes of those  nonzero $\beta_j$ with  weak but nonzero effects.
 Thus,  $S_{1}$ is able to be detected using some existing variable selection techniques,
 while $S_2$ may not be  separated from $S_{3}$  in general  using existing Lasso-type methods.
 A more detailed description can be found in \cite{zhang.Zhang:2014}.
Following the spirit of model parsimony,
covariates in $S_{1}$  are  kept in the model, and some or all covariates in $S_{2}$ are
  left aside with ones in  $S_{3}$. \cite{hansen:2013} has showed using
  simulation studies that such a Lasso estimate
often perform worse than the post selection least square estimate.
To improve the prediction error of  a Lasso-type variable selection approach,
some (modified) post least squares estimators are studied in
\cite{belloni.chernozhukov:2009} and \cite{liu.yu:2013}.
  However, this work still assumes the irrepresentable condition and
those post estimations are only based upon the chosen subset after the Lasso.
 Consequently, the simultaneous weak effects in $S_{2}$ are still ignored.
An ideal strategy should be able to  incorporate
 the joint contribution from covariates in $S_{2}$, even though  a parsimony model without
 including covariates in $S_2$  is adopted.

Let's consider an extreme case where  $S_1$ is a null set and $p$ is fixed.
It has been studied extensively that shrinkage estimators can have uniformly smaller risk
compared to the ordinary maximum likelihood estimators (MLEs) since the discussion papers in
\cite{stein:1956} and \cite{james.stein:1961}.
The relative risk properties of  shrinkage estimators were also investigated  in low-dimensional regression model under a
restricted linear submodel space. See for example,
 \cite{ahmed:2014}, \cite{ahmed.fallahpour:2012},  \cite{ahmed.hossain.doksum:2012},
and many others.

 However, in high-dimensional settings where $p>n$,  {\it a priori} on $S_1$ is not guaranteed,
 not mentioning the existence of an  MLE.
 Thanks to the existing variable selection techniques, an estimated candidate subset $\widehat S_1$
 is selected.
Once $\widehat S_1$ is obtained, the next question we want to ask is: can we
construct a post selection shrinkage estimate to improve the risk of the
post selection least squares estimators?

As we know, ridge regression \citep{marsaglia.styan:1974,frank.friedman:1993}
has been widely used  when the designed matrix is ill-conditioned such
that a regular MLE is not available.
In this paper, we follow the model parsimony spirit and extend shrinkage estimation
to the  high-dimensional data setting using both  ridge penalty and Lasso-type penalty separately.
In particular, we use a ridge penalty  to construct
a data-adaptive post selection shrinkage estimator to improve the
risk of a post selection least squares estimators  based upon a Lasso-type variable selection result.
%The suggested post selection shrinkage estimation  is to shown to outperform the
%%full model weighted ridge estimator in terms of the asymptotic
%quadratic risk.
%More importantly,
%when a selected candidate submodel from
%any Lasso-type method is not consistent with the true model following the model parsimony spirit,
%the proposed shrinkage estimation strategy
%improves the prediction performance significantly.

We summarize our main contributions as follows:
\begin{itemize}
\item[] 1. We propose a post selection shrinkage strategy
to improve the risk of   the Lasso-type estimators in high-dimensional settings.
 This post selection shrinkage strategy is data-adaptive and has some practical applications,
 especially when an ``important'' subset is generated and some covariates with joint weak effects are not selected.
 %The post selection shrinkage strategy  is data-adaptive and
 % is applicable to a host of model selection criteria not in the selection stage.
\item [] 2. We investigate the asymptotical risk of the proposed post selection shrinkage estimators.
 Corresponding asymptotic properties of a predecessor generating those post selection shrinkage estimators
  are also investigated under some regularity conditions.
\end{itemize}

The rest of the paper is organized as follows.
In Section 2, we describe some preliminary model information involved in building a
post selection shrinkage estimator.
%A list of existing variable selection methods are also summarized in this section to obtain
%a submodel.
As a preparation, we introduce some sparsity definitions under certain signal strength levels.
Some existing variable selection results from Lasso are also summarized in this section.
We propose three steps in constructing the shrinkage strategy in Section 3.
In Section 4, we investigate  some  asymptotic
properties of those post selection estimators during three steps in Section 3.
We first investigate some asymptotic normality properties of
the designed weighted ridge estimators under some conditions.
Then we investigate the asymptotic distributional  risks of the linear combination of
the proposed post selection shrinkage estimators.
In Section 5 and 6, we perform some numerical studies using some simulated examples  and a real data application, respectively.
We summarize the paper with some
discussions in the final section. All proofs are given in the Appendix.

\section{Model description and basic notations}\label{sec-modelsetup}
 Let  $\bbeta^*=(\beta^*_{1},\cdots,\beta^*_{p_n})'$
be the true coefficients vector in model \eqref{linear model}.
For any subset $S\subset\{1,\cdots, p_n\}$ with a cardinal value $|S|$,
denote  $\bbeta^*_{S}$ a subvector of $\bbeta^*$ indexed by $S$.
Similar subscripts are used for other submatrices and subvectors.

\subsection{Model sparsity and signal strength}
As introduced in the previous section, the effect of all $p_n$ covariates are characterized into
three categories based upon their signal strength:
important covariates with strong effects in $S_1$,  covariates with no effect in $S_3$,
 and an intermediate group in $S_2$ with joint weak effects.
 In particular, those  {\it signal strength assumptions} of the true model are made explicitly as follows.
\begin{itemize}
\item [] {\bf (A1)}  There exists a positive constant $c_1$, such that $|\beta^*_{j}|>c_1 \sqrt{(\log p_n) /n}$ for $\forall j\in S_{1}$;
\item[] {\bf (A2)} The  parameter vector  $\bbeta^*$ satisfies
that $\|\bbeta^*_{S_2}\|=O(n^\tau)$ for some $0<\tau<1$, where $\|\cdot\|$ is the $\ell_2$ norm;
\item [] {\bf (A3)}  $\beta^*_{j}=0$,  for $\forall j\in S_{3}$.
\end{itemize}

Assumptions (A1-A3) specify those signal strength levels
 in the strong signals set $S_{1}$, weak signals set $S_2$,
 and sparse signal set $S_3$ explicitly.
In particular, (A2) indicates that joint weak
effects in $\bbeta^*_{S_2}$ only grow with $n$ at a certain rate even
though the dimension $p_n$ grows with $n$ fastly. For example,
if (A1) holds for some $c_1>0$ and we let $|\beta_{0j}|<c_1\sqrt{(\log p_n) /n}$ for $j\in S_2$ with $|S_2|<n$,
then $\|\bbeta^*_{S_2}\|<c_1\sqrt{\log(p_n)}<c_1n^\tau$
 even though $p_n=O(\exp(n^{2\tau}))$.

Most existing high-dimensional sparse models investigate the
variable selection consistency by only considering  the existence of the strong signals in (A1)
and sparse signals in (A3). There are very limited work assuming the existence of weak signals in $S_2$.
For example,  besides  a strong signal set in (A1), \cite{zhang.huang:2008}
does not separate $S_2$ and $S_3$ and makes an alternative sparse model assumption,
\begin{itemize}
\item [] {\bf (A2')}  $\sum_{j\notin S_1}|\beta^*_{j}|\le \eta_1$ for some $\eta_1\ge 0$.
\end{itemize}
In their work,   some sufficient conditions are investigated under which the Lasso can select the strong
signal set $S_1$ consistently, following the spirit of the model parsimony.

Our weak and sparse conditions in (A2-A3) are
different from the sparse condition  in  (A2')
where  $S_2$ and $S_3$ are not separated.
If we replace (A2) by (A2') in our signal strength assumptions, then (A2)
becomes  $\|\bbeta^*_{S_2}\|\le \sum_{j\in S_2}|\beta^*_{j}|=\eta_1$,
the joint effects  in $S_2$ being bounded uniformly.
Thus, a true model under (A2') only is less sparse than
one under (A3) only, but more sparse than one in both (A2) and (A3).
On the contrary, a sparse model under both (A2) and (A3) includes the most weak signals,
a sparse model under (A3) only does not have any weak signals, while a sparse model under (A2') only is
in the middle.

\subsection{Parsimonious model selection}
As discussed in Section 1, a penalized least squares (PLS) estimator
 is often adopted to select a parsimonious model for a  high-dimensional regression model in \eqref{linear model},
 \bel{eq-pen-obj}
  \hbbeta_n^{\rm PLS}= \argmin\left\{\sum_{i=1}^n  \left (y_i-\sum_{j=1}^{p_n}x_{ij}\beta_j \right)^2+ \sum_{j=1}^{p_n}
 p_\lm (\beta_j)\right\},
\eel
where $p_\lm (\beta_j)$ is the penalty term on $\beta_j$ with a tuning parameter
controlling the size of selected candidate subset model.
For example, the Lasso takes $p_\lm (\beta_j)=\lm|\beta_j|$ and
the Adaptive Lasso takes $p_\lm (\beta_j)=\lm|\beta_j|/|w_j|$,
where $w_j$ can be taken as an initial estimator of $\beta_j$.
The size of selected subset model depends strongly on the choice of
tuning parameters in \eqref{eq-pen-obj}.
As pointed by \cite{zhang.Zhang:2014},
one turns to ignore weak signals in $S_{2}$ together with $S_3$ and
 select a candidate subset model with only strong signals in $S_1$,
following the model parsimony spirit.

If we let $\widehat S_1\subset\{1,\cdots, p_n\}$ index an active subset from \eqref{eq-pen-obj}, then
a data-adaptive candidate subset model is produced such that
 \bel {h0-beta}
  \hbeta^{\rm PLS}_j=0 \quad {\rm if~and~only~if~}j\notin \widehat S_1.
 \eel

Denote the response vector $\by=(y_1,\cdots, y_n)'$,
all covariates vectors $\bx_j=(x_{1j}, \cdots, x_{nj})'$ for $j=1,\cdots, p_n$,
and the design matrix $\bX=(\bx_1\cdots\bx_{p_n})$.
Without loss of generality, we rearrange the designed vectors
such that $\bX=(\bX_{S_1}| \bX_{S_2}| \bX_{S_3})$,
where $\bX_{S}$ is the submatrix consists of vectors in $S\in\{1,\cdots, p_n\}$.
Below we give two scenarios where $S_2$ cannot be separated from $S_3$.
%In Case 1, a selected candidate subset includes $S_1$ only, i.e.,
%$\widehat S_1=S_1$, while in Case 2, eite

\bigskip
\noindent{\bf Case 1 \citep{chen.donoho.saunders:1998}} %\label{model example 1}
{\it Consider an orthonormal design with $\bX'\bX/n=\bI_n$ and
$\bveps\sim N(0, \bI_n)$.  The PLS with Lasso penalty
provides a soft-threshold estimator with
$\hbeta^{\rm Lasso}_j=\tbeta_j-\lm/n\sgn(\tbeta_j)$ and $0$,
 for $|\tbeta_j|>\lm/n$ and $|\tbeta_j|<\lm/n$,
 respectively. Here  $\tbeta_j=\bx_j'\by/n\sim N(\beta_{0j}, 1/n)$
 is the least squares solution and  $\sgn(\cdot)$ is the sign mapping function. If
 $\min_{j\in S_1}|\beta^*_{j}|>\lm/n>c>\max_{j\in S_2}|\beta^*_{j}|$  for some $c>0$,
 then $P(\widehat S_1=S_1)\to 1$; that is, $P(\hbeta^{\rm PLS}_j=0) \to 1$ for $j\notin S_1$.
 Thus, all weak signals in $S_2$ are omitted together with sparse signals in $S_3$
 using the Lasso approach.
}

\bigskip
%\begin{example} \label{model example 2}
\noindent{\bf Case 2 \citep{wainwright:2009}} {\it Consider a non-singular design such that
 the smallest eigenvalue of $\bX_{S_3^c}'\bX_{S_3^c}/n$ is  larger than some positive constant $c$. If
there exists some $j\in S_2$ such that
$|\beta_{0j}|<|g_j(\lm)|$, where  $g_j(\lm)=\lm\be_j'(\bX_{S_3^c}'\bX_{S_3^c})^{-1}\sgn(\bbeta_{0S_3^c})$
with $\be_j$ being the $j$th column of the identity matrix,
then $P(\{S_1\cup S_2\subseteq \widehat S_1\}\cap \{ S_3\subseteq\widehat S_1^c\})<1$.
 Thus $S_2$ and $S_3$ cannot be separated using the Lasso.
}

%from now on,  we do not separate the
%active set $\widehat S_1$ and truly strong signal set $S_1$,
 % and let $S_1$ unitedly  represent the selected
 %  candidate subset post variable selection.
Some  post selection estimators were proposed
 to improve the prediction performance of the PLS estimator.
For example,  under some regularity conditions,
 \cite{belloni.chernozhukov:2009} and \cite{liu.yu:2013} studied  some post selection least square estimators,
 \bel{eq-restricted-beta}
 \hbbetaRE_{\widehat S_1} = (\bX_{\widehat S_1}'\bX_{\widehat S_1})^{-1}\bX_{\widehat S_1}'\by,
\eel
Here we denote such a post selection least squares estimator
  as  a restricted estimator (RE), written as $\hbbetaRE_{\widehat S_1}$ in this paper.
  For notation's convenience, we omit the phase of ``post selection'' in some future short notations without
causing any confusion.

When $S_1$ and $S_2$ are not separable, we tend to
 select the important subset $\widehat S_1$ such that
$\widehat S_1\subseteq S_1$ for a large enough $\lm$
or $S_1\subset\widehat S_1\subset S_1\cup S_2$
for a smaller $\lm$, following the spirit of model parsimony.
Although  $\hbbetaRE_{\widehat S_1}$  is more estimation efficient
than $\hbbeta^{\rm PLS}_{\widehat S_1}$,
 the prediction risk of   $\hbbetaRE_{\widehat S_1}$ can still be high since
many weak signals in $S_2$ are ignored in $\hbbetaRE_{\widehat S_1}$.
 Our interest is to improve the risk performance of  $\hbbeta^{\rm RE}_{\widehat S_1}$
given in \eqref{eq-restricted-beta} by picking up some information from
$\widehat S_1^c$, a compliment subset of the selected candidate submodel.

 \subsection{Some additional  notations}
Based upon a subset partition $S_1$, $S_2$, $S_3$, we
can partition the true parameters $\bbeta^*=(\bbeta^{*'}_1,\bbeta^{*'}_2,\bbeta^{*'}_3)'$, without loss of generality.
Some notations are shorten for notation's simplicity such that
$\bbeta^{*}_{S_k}=\bbeta^{*}_k$ for $k=1, 2$ and $3$. Similar notations are also adopted for
other subvectors and matrices. For example, after the same  partition,
 the design matrix $\bX=(\bx_1,\cdots, \bx_{p_n})$ can be written as
 $\bX=(\bX_1,\bX_2, \bX_{3})$. We also write $\bX=(\bZ, \bX_3)$
with $\bZ=(\bX_1,\bX_2)$. The row vector of $\bZ$ is denoted as
 $\bz_i=(z_{i1},\cdots, z_{i, p_1+p_2})$ for $1\le i\le n$.

If we denote $p_{k}=|S_k|$ for $1\le k\le 3$ and $p_n=p_1+p_2+p_3$.
 In this paper, we allow $p_n=\sum_{k=1}^3 {p_k}$ to be very large, but restrict
 $q=p_1+p_2\le n$ such that $\bSigma_n=n^{-1}\bZ'\bZ$ is non-singular. If
 $\bSigma_n$ is non-singular, then a generalized inverse matrix is adopted when needed
 in computations.
 Some other  submatrices of $\bSigma_n$ are defined as follows.
\bel{sigma11.2and22.1}
\begin{array}{ll}
&\bSigma_{n11}=\bX_{1}'\bX_{1}/n,
\quad \bSigma_{n22}=\bX_{2}'\bX_{2}/n, \\
&\bSigma_{n12}=\bX_{1}'\bX_{2}/n,
 \quad \bSigma_{n21}=\bX_{2}'\bX_{1}/n,\\
&\bSigma_{n22.1}=n^{-1} \bX_{2}'\bX_{2}-\bX_{2}'\bX_{1}(\bX_{1}'\bX_{1})^{-1}\bX_{1}'\bX_{2}\\
&\bSigma_{n11.2}=n^{-1} \bX_{1}'\bX_{1}-\bX_{1}'\bX_{2}(\bX_{2}'\bX_{2})^{-1}\bX_{2}'\bX_{1}
\end{array}
\eel
 Let $\bU=(\bX_2,\bX_3)$ be a $n\times (p_n-p_1)$ submatrix of $\bX$. Then another partition is written as $\bX=(\bX_1,\bU)$.
Let $\bM_1=\bI_n-\bX_{1}(\bX_{1}'\bX_{1})^{-1}\bX_{1}'$. Then  $\bU'\bM_1\bU$ is a $(p_n-p_1)\times (p_n-p_1)$ dimensional singular matrix
with rank $k_n\ge 0$.
We denote $\varrho_{1n}\le \cdots\le \varrho_{k_n n}$ as all $k_n$ positive eigenvalues of $\bU'\bM_1\bU$.

%Condition   {\bf (B2)} means that  $n^{-1}\bZ'\bZ$ is positive definite
%although $n^{-1}\bX_n'\bX_n$ is not due to $p_n\gg n$. In the asymptotic
%studies in Section \ref{sec-asym}, we will provide more assumptions on
%the decay rate of $n^{-1}\bX_n'\bX_n$.

\section{Post selection shrinkage estimation strategy}\label{sec:HD-SE}
We propose a high-dimensional (HD) post selection shrinkage estimation strategy
 based upon the following  three steps.
\begin{itemize}
 \item []{\it Step 1:}  obtain a data-adaptive candidate subset $\widehat S_1$ following a
  model parsimony spirit, and construct a post selection
 least square estimator $\hbbetaRE_{\widehat S_1}$ using \eqref{eq-restricted-beta};
 \item []{\it Step 2:}  obtain a post select weighted ridge estimator,
         $\hbbeta^{\rm WR}_n=(\hbbeta^{\rm WR}_{\widehat S_1}, \hbbeta^{\rm WR}_{\widehat S_1^c})$, using
  a threshold ridge penalty to be introduced and a submodel $\widehat S_1$ selected from Step 1;
  \item  []{\it Step 3:} obtain a post select shrinkage estimator by  shrinking
   $\hbbeta^{\rm WR}_{\widehat S_1}$  from Step 2 in the direction of  $\hbbetaRE_{\widehat S_1}$ from Step 1.
\end{itemize}
The post selection weighted ridge estimator in Step 2 can handle three scenarios simultaneously: a) the sparsity in  HD data analysis;
b) the strong correlation among covariates; c)  the jointly weak contribution from some covariates.

\begin{remark}\label{Remark 1}
This post selection shrinkage estimation is expected to
improve the risk performance on the selected submodel
 once a variable selection approach in Step 1
 tends to pop out those and only those variables with  strong signal strength,
 that is, $S_1\supset \widehat S_1$ or $S_1\subset\widehat S_1\subset S_1\cup S_2$.
However, if the  model parsimony spirit is not followed and $\lm$ in \eqref{eq-pen-obj} is too small such
that $\widehat S_1\supset S_1\cup S_2$, this post selection shrinkage estimation
is not suggested. Therefore, the effect of the post selection shrinkage estimator
is data adaptive and depends on $\widehat S_1$.
\end{remark}

As a preparation we first construct a post selection weighted ridge estimation
based upon $\widehat S_1$.
 This post selection weight ridge estimation itself is constructed from two steps.

\subsection{Weighted ridge estimation}\label{sec-fe}

Once $\widehat S_1$ is obtained  from Step 1,
we consider to minimize a penalized objective function
with a ridge penalty on coefficients in ${\widehat S_1^c}$,
\bel{part-ridge-obj}
\tbbeta(r_n)=\argmin\{L(\bbeta; \widehat S_1)\}=\argmin\left\{\|\by-\bX_n\bbeta_{n}\|^2+r_n\|\bbeta_{\widehat S_1^c}\|^2\right\}
\eel
where $r_n>0$ is a tuning parameter controlling the penalty effect on $\bbeta_{\widehat S_1}$.
Then  a post selection weighted ridge (WR) estimator $\hbbetaWR(r_n, a_n; \widehat S_1)$ is obtained from,
\bel{eq-threshold}
\hbeta_j^{\rm WR}(r_n, a_n)=\left\{
    \begin{array}{ll}
    \tbeta_j(r_n), & j\in \widehat S_1; \\
    \tbeta_j(r_n) I(\tbeta_j(r_n)>a_n), & j\in \widehat S_1^c,
     \end{array}
     \right.
\eel
where $I(\cdot)$ is the indicator function and $a_n$ is a threshold parameter.
Thus, we obtain estimators of the weak signal subset
\bel{eq-s2hat}
\widehat S_2:=\widehat S_2(\widehat S_1)=\{j\in \widehat S_1^c: \hbetaWR_j(r_n, a_n)\neq 0\}
\eel
and of the sparse subset
\bel{eq-s3hat}
\widehat S_3:=\widehat S_3(\widehat S_1)=(\widehat S_1\cup \widehat S_2)^c.
\eel
Our post selection strategy is only applied when the threshold  parameter $a_n$ satisfies $|\widehat S_2|>2$ and $|\widehat S_3^c|<n$.
 In particular,  we set
\bel{eq-an}
a_n=c_1 n^{-\alpha}, \quad 0<  \le 1/2,~{\rm for~some~}c_1>0.
\eel

\begin{remark}\label{Remark 2}  We call $\hbbetaWR(r_n, a_n)$ a post selection weighted ridge estimator
from two facts: 1) we only penalize parameters in $\bbeta_{\widehat S_1^c}$  instead of
the entire coefficients vector $\bbeta_n$, and
 2) the  threshold step in \eqref{eq-threshold} can be interpreted
as a weighted ridge penalty  $r_{n}(\beta_j^2/w_j^2)$ for $j\in \widehat S_1^c$ in \eqref{part-ridge-obj},
where  $w_j=0$ or $1$ if $j\in \widehat S_3$ or  $j\in \widehat S_2$.
\end{remark}

\begin{remark} \label{Remark 3} Similar to the discussion in Remark \ref{Remark 2},
we can also understand the post selection step into the
weighted ridge estimator, $r_{n}(\beta_j^2/w_j^2)$ with $w_j=\infty$ for
$j\in \widehat S_1$. We do not enforce an additional ridge penalty on $\widehat S_1$
is to reduce some unnecessary biases during the weighted ridge step.
This is different from the post selection threshold regression studied in \cite{zheng.fan.lv:2014},
where the $\ell_2$ penalty is applied on the entire $\bbeta_n$ equally.
\end{remark}

\begin{remark}\label{Remark 4.0} The idea of the weighted ridge regression is connected to the regularization after retention framework proposed in \cite{weng2013regularization}. In that framework, a retention step is conducted to find the important set $\hat S_1$ with large marginal correlation coefficients with the response. Then, a regularization step is conducted by a penalized least square with $L_1$ regularization only on the covariates that are not in $\hat S_1$. Compared to that framework, the current framework focused more on prediction by using the ridge penalty and the estimate $\hat S_1$ is also different.
	
\end{remark}

Notice that, for every selected candidate subset $\widehat S_1$,
$\hbbetaWR_{\widehat S_1}(r_n)$ depends on $r_n$ and $\hbbetaWR_{\widehat S_1^c}(r_n, a_n)$
depends on both $r_n$ and $a_n$.
For  convenience, we omit those tuning parameters and denote above post selection
weighted ridge estimators as $\hbbetaWR_{\widehat S_1}$ and $\hbbetaWR_{\widehat S_1^c}$,
respectively.

\subsection{Post selection shrinkage estimation}\label{sec-se}

Now we are ready to propose a shrinkage estimation based upon
two post selection estimators:    $\hbbetaRE_{S_1}$   and $\hbbetaWR_{S_1}$.

%where we often omit the subscript $a_n$ and denote $\widehat\cM=\widehat\cM_{a_n}$ without causing any confusion.
An initial post selection shrinkage estimator  $\hbbetaSE_{\widehat S_1}$ is defined as
\bel{eq:se}
\begin{array}{ll}
 \hbbetaSE_{\widehat S_1}&= \hbbetaRE_{\widehat S_1} +(\hbbetaWR_{\widehat S_1} -  \hbbetaRE_{\widehat S_1}) (1- (\widehat s_2-2)/ \widehat T_n)\\
                &=\hbbetaWR_{\widehat S_1} -((\widehat s_2-2)/\widehat T_n)(\hbbetaWR_{\widehat S_1} -  \hbbetaRE_{\widehat S_1}),
 \end{array}
\eel
where $\widehat s_2=|\widehat S_2|$  and
 $\widehat T_n$ is given by
\begin{equation}\label{eq:ch1:test:statistic}
\widehat T_n=(\hbbetaWR_{\widehat S_2})'(\bX_{\widehat S_2}'\bM_{\widehat S_1}\bX_{\widehat S_2})\hbbetaWR_{\widehat S_2}/\sigma^2,
\end{equation}
where $\bM_{\widehat S_1}=\bI_n-\bX_{\widehat S_1}(\bX_{\widehat S_1}'\bX_{\widehat S_1})^{-1}\bX_{\widehat S_1}'$.
If $\sigma^2$ is unknown, it is replaced by a consistent estimator
$\hsigma^2$. In the numerical studies,
$\sigma^2$ is replaced by $\hsigma^2=\sum_{i=1}^n (y_i-\bx_i'\hbbetaWR_{\widehat S_2})^2/(n-\widehat s_2)$,
and a generalized inverse is used if $(\bX_{\widehat S_1}'\bX_{\widehat S_1})^{-1}$
is not singular.
%under UPI or AI in \eqref{h0-beta}.
%The HDSE resemble with Stein's estimation. However, we obtain
%\subsection{An improved modification}\label{sec-pse}

  Observing from \eqref{eq:se} and \eqref{eq:ch1:test:statistic},
   signs of two   estimators of $\bbeta_{\widehat S_1}$ can be reversed
   if $\widehat T_n$ is too small such that $\widehat s_2-2>\widehat T_n$.
   It is possible since $\hbbetaWR_{\widehat S_1^c}$ consists of
   nuisance parameters and over-shrinkage can occur
   for a large $r_n$ in the weighted ridge  step.
Thus, we also suggest  to modify \eqref{eq:se} as the following
post selection positive shrinkage estimation (PSE),
\bel{eq-pse}
 \hbbetaPSE_{\widehat S_1}=\hbbetaWR_{\widehat S_1} -([(\widehat s_2-2)/\widehat T_n] \wedge 1) (\hbbetaWR_{\widehat S_1} -  \hbbetaRE_{\widehat S_1}).
\eel

\begin{remark}\label{Remark 4}
Our proposed post selection shrinkage estimation and the classical shrinkage estimation bear some resemblance
 but are different due to two facts: 1) Post selection shrinkage estimation is associated with
  a selected candidate subset and  has some flexibility of adjusting the shrinkage strength data adaptively
since $\hbbetaWR_{\widehat S_1^c}$ depends on tuning parameters $a_n$ and $r_n$;
2)  Post selection shrinkage estimation  uses an initial ridge shrinkage step and
is tailored for the HD settings where multiple covariates tend to be correlated and function jointly.
\end{remark}

\section{Asymptotic properties}\label{sec-asym}

In order to investigate some asymptotic properties of the proposed
post selection estimators,  we
first make following assumptions on the random error,  $\bU'\bM_1\bU$ and
 the model sparity.
One can review some  notations at the end of Section \ref{sec-modelsetup}.
\begin{itemize}
\item[] {\bf (B1)}  The random error $\veps_i\sim N(0, \sigma^2)$.
\item[] {\bf (B2)} $\varrho_{1n}^{-1}=O(n^{-\eta}),$ where $\tau<\eta\le 1$ for $\tau$ in (A2).
\item[] {\bf (B3)} $\log(p_n)=O(n^\nu)$ for $0<\nu<1$.
%\begin{itemize}
%\item[] {\bf (B4):}  $\tau<1/2$ and  $r_n=o(n^{1/2-\tau})$.
\item[] {\bf (B4)} There exist a positive definite matrix $\bSigma$ such that $\lim_{n\to \infty} \bSigma_n=\bSigma$,
      where eigenvalues of $\bSigma$ satisfy $0<\rho_1<\rho_\bSigma<\rho_2<\infty$.
\end{itemize}
%\textcolor{red}{we may replace $\bC_{-3}$ by a different notation}
Here condition   (B1) can be relaxed to a symmetric distribution
with some finite moments. To simplify our theoretical investigations  and handle
the ultra high-dimensionality, we only restrict our studies to normal random error in this paper.
Condition (B2) guarantees that the positive eigenvalues of the redundant $\bU=\bX_{S_1^c}$
cannot be too small with a rate associated with the  weak signals strength in $S_2$.
Condition (B3) permits the ultra-high-dimensionality such that the number of variables
can grow with sample size at an almost exponential rate.
Condition in (B4) is the regularity condition for $\bX_{S_3^c}$.
This condition is made in order to obtain the asymptotic normality the
 weighted ridge estimator.

\subsection{Asymptotic properties of the weighted ridge estimator}\label{sec-asym-norm}

We have the following  asymptotic  properties of the
weighted ridge estimator $\hbbetaWR_n$.
\begin{theorem} \label{thm-variable consistency}
Suppose the sparse model in \eqref{linear model} satisfies signal strength assumptions in
 (A1-A3) and  model assumptions in (B1-B3).
 If we choose $r_n=c_2 a_n^{-2}(\log\log n)^3 \log(n\vee p_{n})$ for some constant $c_2>0$ and
 $a_n$ defined in \eqref{eq-an} with $\alpha< (\eta-\nu-\tau)/3$, then
 $\widehat S_2$ in \eqref{eq-s2hat} satisfies
\bel{eq-consistency}
P(\widehat S_2=S_2|\widehat S_1=S_1)\ge 1-(n\vee p_{n})^{-t}~{\rm for~some~constant~} t>0,
%P(\widehat S_2=S_2|\widehat S_1=S_1)\ge 1-\exp\{ (1-c_3\log\log n) n^{\nu}\}~{\rm for~constant~}c_3>0,
\eel
where $\tau$, $\eta$, and $\nu$ are defined in (A2), (B2) and (B3), respectively.
\end{theorem}
Theorem \ref{thm-variable consistency} is similar to the variable selection result in
 \cite{shao.deng:2012}.
  We postpone the detailed proof to the Appendix.
It  tells us that
the weighted ridge estimator $\hbbetaWR_{S_1^c}$ is able to
single out the sparse set $S_{3}$ with a large probability,
if $S_1$ is pre-selected  in advance such that $P(\widehat S_1=S_1)=1$.
%This seems not to be very {\it practical} since $\widehat S_1$
For example, \cite{zhang.huang:2008} argued that $S_1$ can be recovered with
a large probability under the sparse Riesz condition (SRC) with rank $p_1$. Here
a design matrix $\bX$ satisfies the SRC with rank $q$ and spectrum bounds $0<c_*<c^*<\infty$ if
\bel{eq-src}
c_*\le \dfrac{\|\bX_S\bv\|^2}{\|\bv\|^2}\le c^*\quad \forall S~{\rm with~}|S|=q~{\rm and~} \bv\in \cR^q.
\eel
\begin{lemma}\label{lemma1}
Consider the Lasso solution for  linear model \eqref{linear model} with $\veps_i\sim N(0, \sigma^2)$.
Suppose  (A1), (B1) are satisfied and the sparse condition  (A2') holds for some $0<\eta_1<O(p_{1}\sqrt{\log(p_n)/n})$, and
  the design matrix $\bX$ satisfies the SRC with rank $p_1$ in \eqref{eq-src}.
 Then $\widehat S_1$ generated from a PLS with the Lasso penalty in \eqref{eq-pen-obj} satisfies
$$
\displaystyle\lim_{n\to \infty} P\left(\{S_1\subset \widehat S_1\}\cap\left\{\sum_{j\in S_1}|\beta_j^*|I(\hbeta^{\rm PLS}_j=0)=0\right\} \right )
=\displaystyle\lim_{n\to \infty} P(S_1=\widehat S_1)=1.
%\ge 2-\exp\left\{\dfrac{2p_n}{p_n^{(1+c_0)}}\right\}-\dfrac{2}{p^{1+c_0}}\approx 1~{\rm for~some~} c_0>0.
$$
\end{lemma}
Lemma \ref{lemma1} is a direct result from Theorem 2 in \cite{zhang.huang:2008}.
Here the tuning parameter in \eqref{eq-pen-obj} is chosen such that
 $\lm\ge 2\sigma\sqrt{2(1+c_0)c^*n\log(p_n)}$. Lemma \ref{lemma1} indicates
that those and only those strong signals in $S_1$ are included in $\widehat S_1$ while using the Lasso
under sufficient conditions.

In Lemma \ref{lemma1},  we have $\sum_{j\notin S_1}|\beta_j^*|<\eta$. The signal of each individual coefficient
is trivial if such a joint effect is uniformly distributed on $\ge p_n-n$ coefficients when $p_n\gg n$. However,
if this  joint effect is only distributed on a much smaller number of coefficients, each individual effect may not
be negligible. In particular, if we let both (A2') and (A3) hold, then  $\sum_{j\in S_2}|\beta_j^*|<\eta$. Thus
(A2) also holds. Combing Lemma 1 and Theorem 1, we have following result directly.

\begin{corollary}\label{cor1}
Suppose all conditions in both Lemma \ref{lemma1} and Theorem \ref{thm-variable consistency} hold. Then we have
\bel{eq-consistency2}
\displaystyle\lim_{n\to \infty} P\left(\{\widehat S_2=S_2\} \cap \{\widehat S_1=S_1\}\right)=1.
\eel
\end{corollary}
Corollary \ref{cor1} indicates that $\widehat S_3=S_3$ is
  able to be recovered  if an additional weighted ridges step is used post the Lasso
  under some sufficient conditions.
We skip the proof since this is a direct result from Lemma 1 and Theorem 1.
%A short proof of Corollary 1 is provided in the Appendix.

However, Corollary \ref{cor1} still requires a SRC condition.
Although $P(\widehat S_1=S_1)=1$ may not be guaranteed when a SRC condition is not satisfied,
we may have
\bel{eq-new1}
 P\left(\{S_1\subset \widehat S_1 \subset S_1\cup S_2\} \right )\to 1.
 \eel
 Thus, we  have similar but weaker result.
\begin{corollary}\label{cor2}
Suppose all conditions in Theorem \ref{thm-variable consistency} hold and
$\widehat S_1$ satisfies \eqref{eq-new1}. Then we have
\bel{eq-consistency2}
\displaystyle\lim_{n\to \infty} P\left(\{\widehat S_2=\widehat S_1^c\cap S_2\}\right)=1.
\eel
\end{corollary}
Corollary \ref{cor2} can be interpreted by
treating $\widehat S_1$ as a new $S_1$ and $\widehat S_1^c\cap S_2$ as a new $S_2$.

The asymptotic properties in Theorem  1 and  its derivatives in Corollary \ref{cor1} and \ref{cor2}
 are important for
establishing the efficiency of $\hbbetaWR_{\widehat S_1}$ and  $\hbbetaWR_{\widehat S_2}$.

\begin{theorem}{\label{thm-normality}}%
Let $s_n^2=\sigma^2\bd_n'\bSigma_n^{-1} \bd_n$ for any $(p_{1n}+p_{2n})\times 1$ vector $\bd_n$ satisfying
$\|\bd_n\|\le 1$. Suppose  assumptions  (B1-B4) hold. Consider
a sparse model with signal strength under  (A1), (A3), and (A2) with $0<\tau<1/2$.
Suppose a pre-selected model such as $S_1 \subset \widehat S_1 \subset S_1\cup S_2 $ is obtained  with probability 1.
If we choose $r_n$ as in Theorem \ref{thm-variable consistency}  with $\alpha<\{(\eta-\nu-\tau)/3, 1/4-\tau/2\}$, then
 we have the asymptotic normality,
\bel{eq-betanormality}
n^{1/2} s_n^{-1}\bd_n' (\hbbetaWR_{S_3^c}-\bbeta^*_{S_3^c})
~\underrightarrow{{\rm d}}N(0, 1).
\eel
\end{theorem}
Theorem 2 studies the  asymptotic normality of the weighted ridge estimator,  $\hbbeta_{S_3^c}$.
In addition,  $\hbbeta_{S_3^c}$ has the same estimation efficiency as
 one  from a restricted least square estimator as if $\bbeta_{S_3}=0$ is given as {\it a priori}.
However, the result holds if $\|\bbeta^*_{S_2}\|=o(n^{1/2})$ and
$r_n$ is chosen  appropriately.
More importantly, the strong signal set $S_1$ is detected with a large probability in advance.
This can be guaranteed under Lemma \ref{lemma1}.

\subsection{Asymptotic distributional risk analysis} \label{sec-asym-biasandrisk}
In this section, we  provide the relative performance of
 the post selection shrinkage estimation regarding
 the  asymptotic distribution
risk (ADR) introduced in \cite{saleh:2006}.
 For simplicity and notation's convenience,
we focus on the ADR analysis by assuming $\widehat S_1=S_1$ following
the spirit of model parsimony.
If $S_1\subset \widehat S_1\subset S_1\cup S_2$, the similar analysis
can be done by redefining $(S_1, S_2)=(\widehat S_1, \widehat S_1 \cap S_2)$,
as discussed in Section \ref{sec-asym-norm}.
Together with results in Theorem \ref{thm-variable consistency} such that
 $P(\widehat S_3=S_3)\to 1$, $S_3$ is also  removed from the post
 selection  shrinkage estimator with a large probability.
 Thus the risk analysis in this section will be conducted
by assuming both $S_1$ and $S_3$ are known in advance.

\begin{definition}\label{definition 1}
For any estimator $\bbeta_{1n}^\diamond$ and  $p_{1n}-$dimensional vector, $\bd_{1n}$,
satisfying $\|\bd_{1n}\|\le 1$, the ADR of $\bd_{1n}'\bbeta_{1n}^\diamond$
is
\bel{eq-risk}
  {\rm ADR}(\bd_{1n}'\bbeta_{1n}^\diamond )= \lim_{n\to\infty} E\{[n^{1/2}s_{1n}^{-1}\bd_{1n}'(\bbeta_{1n}^\diamond- \bbeta^*_{1})]^2\},
\eel
where  $s_{1n}^2=\sigma^2\bd_{1n}'\bSigma_{n11.2}^{-1} \bd_{1n}$ with $\bSigma_{n11.2}$ defined in \eqref{sigma11.2and22.1}.
\end{definition}

We will provide some analytic  expressions of ADRs under
 specific weak coefficients
 in (A2'').
In particular,
\begin{itemize}%\label{eq:local:alternative:shrinkage:lasso}
\item[]{\bf (A2'')} $\beta^*_{j}=\delta_j/\sqrt{n} ~{\rm for~}j\in S_2$, where $|\delta_j|<\delta_{\max}$ for some $\delta_{\max}>0$.
\end{itemize}
Denote $\bdelta= (\delta_1,  \cdots, \delta_{p_{2n}})'  \in\cR^{p_{2n}}$.
Then  $\Delta_n=\bdelta'\bSigma_{n22.1}\bdelta\le \rho_2 p_{2n}\delta_{\max}$,
where  $\rho_2$ is defined in (B4).

Define
\bel{eq-Delta-d1n}
\Delta_{\bd_{1n}}=\frac{\bd_{1n}'(\bSigma_{n11}^{-1}\bSigma_{n12}\bdelta\bdelta'\bSigma_{n21}\bSigma_{n11}^{-1})\bd_{1n}}
{\bd_{1n}'(\bSigma_{n11}^{-1}\bSigma_{n12}\bSigma_{n22.1}^{-1}\bSigma_{n21}\bSigma_{n11}^{-1})\bd_{1n}}.
\eel
We obtain the following results on the expression of ADRs of post selection shrinkage estimators.

\begin{theorem}{\label{thm-aqr}}
%Suppose $\bbeta_{0}$ satisfies  $K_n$ in \eqref{eq:local:alternative:shrinkage:lasso}.
Let $\bd_{1n}$ be any $p_{1n}-$dimensional vector
satisfying $0<\|\bd_{1n}\|\le 1$ and
$s_{1n}^2=\sigma^2\bd_{1n}'\bSigma_{n11.2}^{-1} \bd_{1n}$.
Suppose all assumptions in Theorem \ref{thm-normality} hold except that
(A2) is replaced by (A2''). Then  we have
\begin{subequations}
\begin{align}
{\rm ADR}(\bd_{1n}'\hbbetaWR_{1n} )&=1, \label{aqr-ue}\\
{\rm ADR}(\bd_{1n}'\hbbetaRE_{1n} )&=1-(1-c)(1-\Delta_{\bd_{1n}}), \label{aqr-re}\\
{\rm ADR}(\bd_{1n}'\hbbetaSE_{1n} )&=1-E[g_1(\bz_2+\bdelta)], \label{aqr-se}\\
{\rm ADR}(\bd_{1n}'\hbbetaPSE_{1n} )&=1-E[g_2(\bz_2+\bdelta)]. \label{aqr-pse}
\end{align}
\end{subequations}
Here
$c=\lim_{n\to \infty} \bd_{1n}'\bSigma_{n11}^{-1} \bd_{1n}/(\bd_{1n}'\bSigma_{n11.2}^{-1} \bd_{1n})\le 1,$
%is $0\le c\le 1$,% $\Delta_{\bd_{1n}}$ is defined in \eqref{eq-Delta-d1n},
 $\bz_2$ satisfies that $ s_{2n}^{-1} \bd_{2n}'\bz_2 \to  N(0,1)$
with $\bd_{2n}=\sigma^2\bSigma_{n21}\bSigma_{n11}^{-1}\bd_{1n}$ and $s_{2n}^2=\bd_{2n}'\bSigma_{n22.1}^{-1}\bd_{2n}$.
In addition,
 \bel{eq-g1}
g_1(\bx)=\lim_{n\to\infty}(1-c)
      \dfrac{p_{2n}-2}{\bx'\bSigma_{n22.1}\bx}
      \left[ 2- \dfrac{\bx' ((p_{2n}+2)\bd_{2n}\bd_{2n}')\bx}{s_{2n}^{2}\bx'\bSigma_{n22.1}\bx}\right],
\eel
  and
 \bel{eq-g2}
 \begin{array}{ll}
 g_2(\bx)&=\lim_{n\to\infty}\dfrac{p_{2n}-2}{\bx'\bSigma_{n22.1}\bx}\left[(1-c)\left(2-\dfrac{\bx' ((p_{2n}+2)\bd_{2n}\bd_{2n}')\bx}{s_{2n}^{2}\bx'\bSigma_{n22.1}\bx}\right)\right]
               I(\bx'\bSigma_{n22.1}\bx\ge p_{2n}-2)\\
       &\quad +\lim_{n\to\infty}
               [(2-s_{2n}^{-2}\bx'\bd_{2n}\bd_{2n}'\bx)(1-c)]I(\bx'\bSigma_{n22.1}\bx \le p_{2n}-2),
\end{array}
 \eel
with $I(\cdot)$ being an indicator function.
\end{theorem}
 Theorem 3 lists the analytic expressions of
 the asymptotic risk of all above estimators.
 From Theorem \ref{thm-aqr}, we can obtain the following risk comparisons.

\begin{corollary}{\label{cor3}}
Under assumptions in Theorem \ref{thm-aqr}, we have
%\bel{eq-compare-ineq}
\\
(i)
$
{\rm ADR}(\bd_{1n}'\hbbetaPSE_{1n} )\le {\rm ADR}(\bd_{1n}'\hbbetaSE_{1n} )\le {\rm ADR}(\bd_{1n}'\hbbetaWR_{1n} )
$ holds for $0<\|\bdelta\|^2\le 1$;
\\(ii) Inequalities in (i) also hold for $\|\bdelta\|^2\le 1+\iota$ for some $\iota>0$
if $\Delta_n=\iota p_{2n}$.
\\
(iii)
%$
%{\rm ADR}(\bd_{1n}'\hbbetaRE_{1n} )
%\le  {\rm ADR}(\bd_{1n}'\hbbetaPSE_{1n} )< {\rm ADR}(\bd_{1n}'\hbbetaWR_{1n} )\}
%$ holds for  $\bdelta=0$, where
%the ``='' holds when $p_{2n}\to \infty$.
If  $\|\bdelta\|=o(1)$, then
${\rm ADR}(\bd_{1n}'\hbbetaRE_{1n} )
\le  {\rm ADR}(\bd_{1n}'\hbbetaPSE_{1n} )< {\rm ADR}(\bd_{1n}'\hbbetaWR_{1n} )\}
$ holds for  $\bdelta=0$, where
the ``='' holds when $p_{2n}\to \infty$.
\end{corollary}

Corollary \ref{cor3} indicates that the performance of the post selection shrinkage estimator
is closely related to the post select least squares estimator.
On  one hand, if  $\widehat S_1\subset S_1\cup S_2$ and $(S_1\cup S_2)\cap \widehat S_1^c$ is large,
 then the post selection PSE tends to  dominate the RE.
Thus, the post selection PSE can improve the performance of  the post selection least squares estimators
in \cite{belloni.chernozhukov:2009} and \cite{liu.yu:2013}, especially
 when $p_{n}\gg n$ and an under-fitted submodel is selected by a
 large  penalty parameter.
On the other hand, if a variable selection approach almost generates the right submodel
and  $\|\bdelta\|=o(1)$, i.e., $\lim_{n\to \infty}\widehat S_1= S_1\cup S_2$,  then a post selection LSE
 ($\hbbetaRE_{1n}$) is the most efficient one compared with
all other post selection estimates.
%We postpone the proof of both Theorem \ref{thm-aqr} and Corollary \ref{cor3} to the Appendix.

\begin{remark}
In the high-dimensional setting where $p\gg n$,
we do need to assume the true model to be  sparse in
the sense that most coefficients goes to $0$ when $n\to \infty$. However,
we still permit  some
$\beta_j$ to be small, but not exactly $0$.
Such covariates with a small amount of influence on the response variable
 are often ignored incorrectly in
HD variable selection methods. If we
borrow  information from those covariates
using the proposed shrinkage methods,
the prediction performance based on selected submodel can be improved substantially.
\end{remark}

\section{Simulation studies }
In this section, we use some simulation studies to examine the quadratic
risk  performance of the proposed estimators.
 Our simulation is based on the linear regression model
 in \eqref{linear model}.
 % with a large number of explanatory variables.

{\it True Model Setting.} In all experiments,  $\veps_i$'s are simulated from
independent and identically distributed standard normal random variables,
$x_{is}=(\xi_{(is)}^1)^2+\xi_{(is)}^2$, where
 $\xi_{(is)}^1$ and $\xi_{(is)}^2$, $i=1,\cdots, n$, $s=1,\cdots, p_n$
 are also independent copies of standard normal distribution.
  In all  experiments, we let $n=200$ and $p_n=n^\tau$ for different sample size $n$,
where $\tau$ changes from $1$ to $1.2$ with an increment of $0.02$.
%We consider the true model $\bbeta_0$ with $\bbeta_{30}=\bzero_{p_{3n}}$.
%We consider the UPI of
%In all simulation studies, we consider
 % $\bbeta_{20}=\bzero_{p_{2n}}$ and $\bbeta_{30}=\bzero_{p_{3n}}$.
Three different coefficient configurations are considered as follows.
%\begin{theorem}{Example.}{}
%We consider three different cases in the true model.
\begin{itemize}
\item[]Case 1: $\bbeta^*= (5, 5,
5,\underbrace{ 0.5,\cdots,  0.5}_{10},\bzero_{p_{3n}}')'$;
\item[]Case 2:
$\bbeta^*= (10,10,
10,\underbrace{ 0.1,\cdots,  0.1}_{50})',\bzero_{p_{3n}}')'$;
\item[]Case 3: $\bbeta^*= (10, 10, 10,\underbrace{ 0.1,\cdots,
 0.1}_{p_{2n}},\bzero_{20}')'$.
\end{itemize}
All nonzero coefficients are randomly assigned to be either positive or negative.
Both zero and weak signals coexist in the above three settings.
In Case 1, most covariates are noises.
Compared with Case 1,
 the weak signals become weaker and
 the strong signals become stronger in Case 2.
 In addition, the number of weak signals is larger but also fixed.
 In Case 3, only $p_{3n}=20$ zero signals, large amount of weak signals contribute simultaneously,
 and the number of weak signals grows with the number of covariates
  such that $p_{2n}\gg n$. Notice that the signal
 strength setting in this case is
 different from that considered in our post selection shrinkage analysis,
 where $p_{2n}<n$ and $p_{3n}\gg n$.

{\it Subset selection.}
Since the Adaptive Lasso, SCAD and MCP perform closely under certain conditions,
we only adopt the Adaptive Lasso and Lasso in selecting a subset before
the post selection shrinkage strategy is applied.
 All tuning parameters in variable selection approaches are chosen  using the BIC.
% However, we only show results for $p_{1n}=4$ in Table~3.

%\end{theorem}
{\it Tuning parameters and simulation Setting.}
As we know, $a_n$ and $r_n$ are two important tuning parameters
affecting $\widehat S_2$ and $\widehat S_3$.
We choose those two tuning parameters based upon the asymptotic
investigations in Theorem \ref{thm-normality} for all our numerical studies.
In particular, the post selection PSEs are computed for
$r_n=c_2 a_n^{-2}(\log\log n)^3 \log(n\vee p_{n})$ with
 $a_n=c_1 n^{-1/8}$. Corresponding coefficients $c_1$ and $c_2$ are chosen using cross validation.
%The above choices of $r_n$ and $a_n$ are based upon
% 1) $p_n=n^\alpha$ and
%2)  $\tau<1/2$ and  $r_n=o(n^{1/2-\tau})$ in (B4). In other words,  we adopt $n^{0.125}\le r_n\le n^{0.225}$
%in the simulation so that condition (B4) is satisfied.

{\it Evaluation.} Each design is repeated 1000 times, as a further increase in the
number of realizations did not significantly change the result.
 Let $\bbeta_{1n}^\diamond$ be either $\hbbetaPSE_{1n}$ or $\hbbetaRE_{1n}$
 after variable selection.
 The performance of  $\bbeta_{1n}^\diamond$ is
evaluated  by the relative mean squared error (RMSE)
criterion with respect to $\hbbetaWR_{1n}$
as follows:
\bel{eq-rmse}
{\rm RMSE} (\bbeta_{1n}^\diamond)=\frac{E\|\hbbetaWR_{1n}-\bbeta^*_{1}\|^2}{E\|\bbeta_{1n}^\diamond-\bbeta^*_{1}\|^2}.
\eel
Therefore, RMSE$(\bbeta_{1n}^\diamond)>1$ means the superiority of $\bbeta_{1n}^\diamond$ over $\hbbetaWR_{1n}$.

\begin{figure}[tp]
  \caption{RMSEs of post selection PSEs compared with one from Lasso or Adaptive Lasso (ALasso) from simulation examples in
  Case 1-3.
  The top (a or a'), middle (b or b'), bottom (c or c')panels are for (a-c), respectively.
  The left (a-c) and right panels (a'-c') are comparisons
  when the candidate submodels are chosen from the  Lasso and Adaptive Lasso methods, respectively.}\label{fig3abc}
\centering
 $$
   \scalebox{0.4} [0.40]{\includegraphics{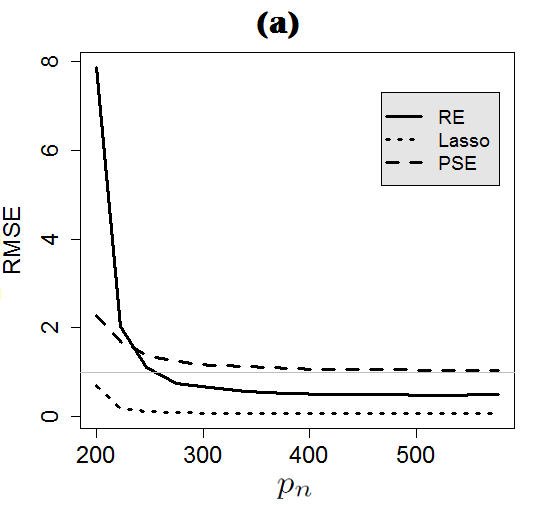}}
     \scalebox{0.4}[0.40]{\includegraphics{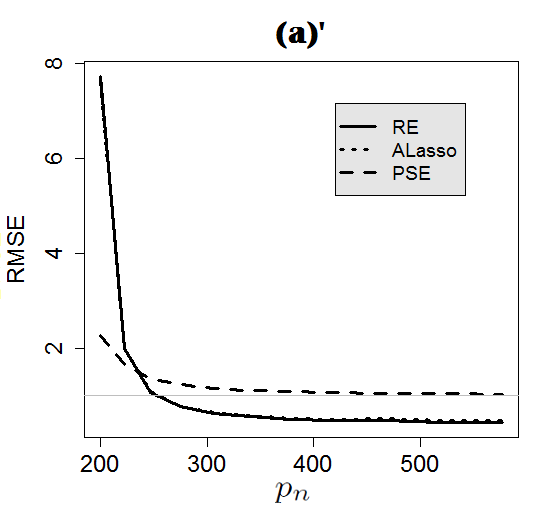}}
  $$
   $$
   \scalebox{0.40}[0.40]{\includegraphics{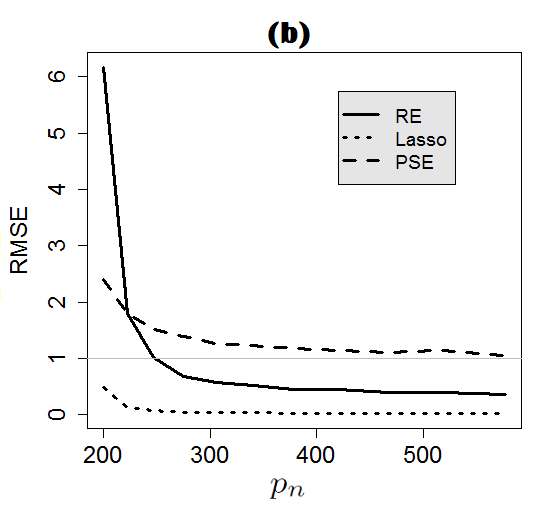}}
      \scalebox{0.40}[0.40]{\includegraphics{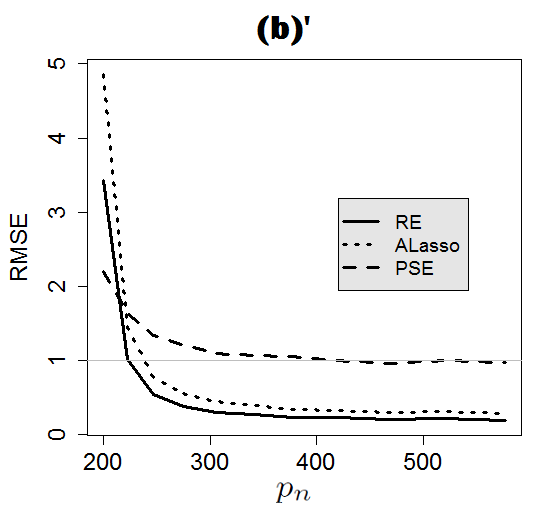}}
  $$
  $$
   \scalebox{0.35}[0.30]{\includegraphics{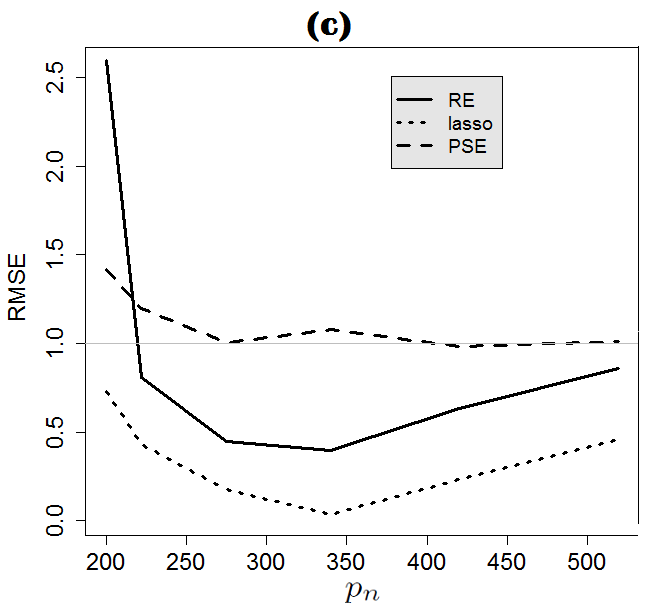}}
   \scalebox{0.40}[0.40]{\includegraphics{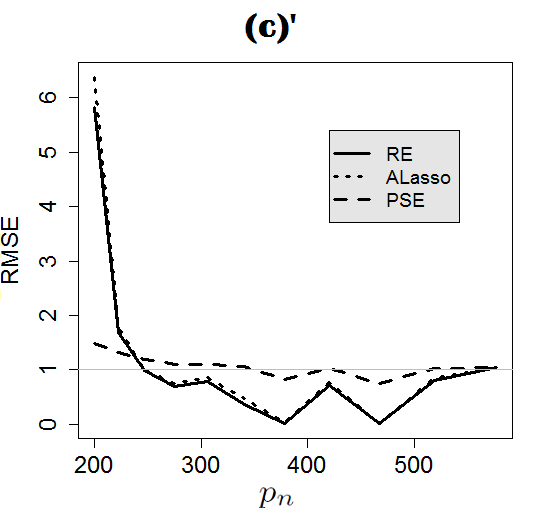}}
  $$
\end{figure}

{\it Result:}  We plot the mean RMSEs from $1000$ iterations
  along $p_n$  in Figure \ref{fig3abc}.
 Some selected results are also reported in Table \ref{table3abc}.
 To check the behavior of Lasso or Adaptive Lasso for subset selection, we also
 report the average number of selected important
 covariates  as $|\widehat S_1|$ in Table \ref{table3abc}.
 It is not surprising to see that
  RE post the Adaptive Lasso is comparable
    to the Adaptive Lasso itself, while
  RE post the Lasso behaves much better than Lasso
  \citep{belloni.chernozhukov:2009,liu.yu:2013}.
We summarize the simulation results as follows:
 \begin{itemize}
 \item Figure \ref{fig3abc} (a')-(c') list results when the Adaptive Lasso is used to generate
  the submodel.
  (1) When $p_n$ is closer to $n$,  both post selection RE and Adaptive Lasso performs better than the post selection PSE
         and WR (RMSE$>$1). (2) When $p_n$ grows bigger, both RE and Adaptive Lasso become worse
  than the post selection WR (RMSE$<$1). However, the post selection PSE  still performs  better than the post selection WR. Therefore,
  The post selection PSE  provides a protection of the Adaptive Lasso in the case that Adaptive Lasso
  loses its efficiency.
  \item Figure \ref{fig3abc} (a)-(c) list results when the Lasso is used to generate
  the submodel.
The advantage of the post selection PSE over the Lasso is more obvious than the above.
This is due to the fact that the Adaptive Lasso tends to produce a more efficient estimator
than the Lasso does.
  \item  When  $p_n$ grows, the post selection PSE is much more robust and at least as good as the weighted ridge estimator
(RMSE is approaching to 1).
When $p_n$ grows bigger, the improvement of the post selection PSE from Adaptive Lasso or Lasso
  become more obvious. See Table \ref{table3abc}.
\item In Case 3,  the post selection PSE may lose its superiority to the post selection RE and Adaptive Lasso
especially when   $p_n$ grows with $n$ fast. One explanation is that
  the  selected model size varies dramatically since the number of
  weak coefficients grows.
 However, if we still follow the model parsimony spirit
  and decide to use an aggressive tuning parameter to get
  a relatively consistent submodel size $\widehat S_1$, the superiority of post selection PSEs follows the
  same pattern as  in Cases 1 and 2.
 \end{itemize}

\begin{table}[h]
\begin{center}
 \caption{Simulated RMSEs from simulation examples in Case 1-3.}\label{table3abc}
\begin{tabular}{llrrrrrrrrr}\\ \hline\hline
%\toprule
\rule{0pt}{3ex}
  && \multicolumn{4}{c}{Lasso} && \multicolumn{4}{c}{Adaptive Lasso}\\
Case & $p_n$ & DF& $\hbbeta_{1n}^{\textrm{Lasso}}$ &  $\hbbetaTRE_{1n}$ & $\hbbetaPSE_{1n}$
             && DF&  $\hbbeta_{1n}^{\textrm{ALasso}}$ &  $\hbbetaTRE_{1n}$ & $\hbbetaPSE_{1n}$   \\
\hline
(a) & 200& 10.920 &0.690 &7.880 &2.285&&10.537 & 7.611 &7.739 &2.269\\
    & 222& 10.785 &0.190 &2.035 &1.680&&10.434  &2.001& 1.991& 1.667\\
    & 275& 10.655 &0.082 &0.744 &1.231&&10.250  &0.783& 0.773 &1.242\\
    & 340& 10.491 &0.066 &0.574 &1.126& &10.137 & 0.585 &0.558 &1.114\\
    & 420&10.416 &0.061 &0.485 &1.061&&9.906 & 0.514 &0.491 &1.062\\
    & 519& 10.293 &0.063 &0.476 &1.047&& 9.781  &0.480 &0.446 &1.042\\
\\
(b) &200 & 3.112& 0.491 &6.169 &2.409 && 3.170&  4.859 &3.431 &2.199\\
&222 &3.078 &0.137 &1.790 &1.807     &&3.149 & 1.447 &1.012 &1.640\\
&275 &3.041 &0.048 &0.684 &1.393    &&3.083  &0.561 &0.384 &1.205\\
&340 &3.036 &0.035 &0.517 &1.222    &&3.051  &0.395 &0.270 &1.066\\
&420 &3.000 &0.029 &0.442 &1.138    &&3.025  &0.335 &0.233 &1.003\\
&519 &3.000 &0.023 &0.388 &1.140    &&3.000  &0.312 &0.217 &0.998\\
        \\
   (c)& 200 &4.020 &0.730 &2.594 &1.420 &&7.379 & 6.380 &5.815 &1.491\\
    & 222&6.109 &0.430 &0.809 &1.200 && 10.005  &1.778 &1.684 &1.310\\
    & 275 &5.277 &0.176 &0.449 &1.007   & &8.159  &0.747 &0.687 &1.092\\
    & 340&3.046 &0.034 &0.396 &1.077 & & 3.783  &0.476 &0.361 &1.070\\
    & 420 &5.325 &0.231 &0.633 &0.984 & &7.390  &0.762 &0.710 &1.025\\
    & 519&7.213 &0.461 &0.860 &1.014 & &9.114  &0.844 &0.804 &1.020\\
 \\ \hline\hline
\end{tabular}
\end{center}
\end{table}

\section{Real data example}
In this section, we apply the proposed post selection shrinkage strategy
to the growth data  for the years 1960-1985 \citep{barro.lee.1994}.
Table  \ref{data-table1} lists the detailed descriptions
of the dependent variable
and $45$ covariates
related to education and its interaction with ${\rm lgdp60}_i$,
  market efficiency, political
stability, market openness and demographic characteristics.

The growth regression model has been applied to test
the negative relationship between the long-run growth rate and the initial GDP given other
covariates. See \cite{barro.sala:1995} and \cite{durlauf.johnson.temple:2005} for literature reviews.
Very recently, \cite{lee.seo.shin:2014} took into account
 the possible discrepancy among the above negative relationship
 using a growth regression model with threshold. In particular,
 they consider a threshold variable in the following regression model,
 \bel{reg-model-data}
 {\rm gr}_i=\beta_0+\beta_1 {lgdp60}_i +\bz_i'\bbeta_2+
 I(Q_i<\tau)(\delta_0+\delta_1 {lgdp60}_i +\bz_i'\bdelta_2)+\veps_i,
 \eel
 where ${\rm gr}_i$ is the annualized GDP growth rate of country $i$ from 1960 to 1985,
 ${\rm lgdp60}_i$ is the log GDP in 1960,
 $\bz_i$ includes all 45 covariates listed in Table  \ref{data-table1},
and $Q_i$ is a threshold variable, where we use the initial GDP in 1960.
 Since the estimation of the threshold parameter $\tau$ is not our target, we
  consider 5 different $\tau$'s in our analysis:  $1655,    2073,  2898,     3268,    6030$.
  Among them, $\tau=2898$ is a threshold value suggested by \cite{lee.seo.shin:2014},
 the other 4 threshold values are $k$th percentiles for $k=60, 70, 80, 90$, respectively.
After removing all missing data,
 each setting includes $n=82$ observations  and $p=90$ covariates besides two intercepts.
% We omit the intercept after centralizing the data for convenience.

Before applying the post selection shrinkage strategy, we first obtain candidate subsets from  two variable
selection techniques:   Lasso and Adaptive Lasso, respectively.
All tuning parameters are selected from 5-fold cross validation.
 In Table \ref{data-table2}, we list the numbers of selected important variables,
 $\widehat s_1=|\widehat S_1|$, also the sizes of candidate submodels,
under 5 different $\tau$'s.
 In Table \ref{data-table3}, we list the frequency of each variable
 being selected among all 5 settings.
  We observe that Lasso and Adaptive Lasso variable
 selection results are quite close for this data set. However,
 the selected candidate subset model  can be quite different
among all five different $\tau$'s.

 After the variable selection, post selection PSE is  applied based upon the selected candidate subsets in all settings.
Table  \ref{data-table4} and
\ref{data-table5} give the
the estimation results for $\tau=2898$ and $\tau=1655$, where both candidate subsets are selected by
the Adaptive Lasso.  We omit  results under other settings to save the space.

\begin{figure}[tp]
\centering
 $$
   \scalebox{0.4}[0.4]{\includegraphics{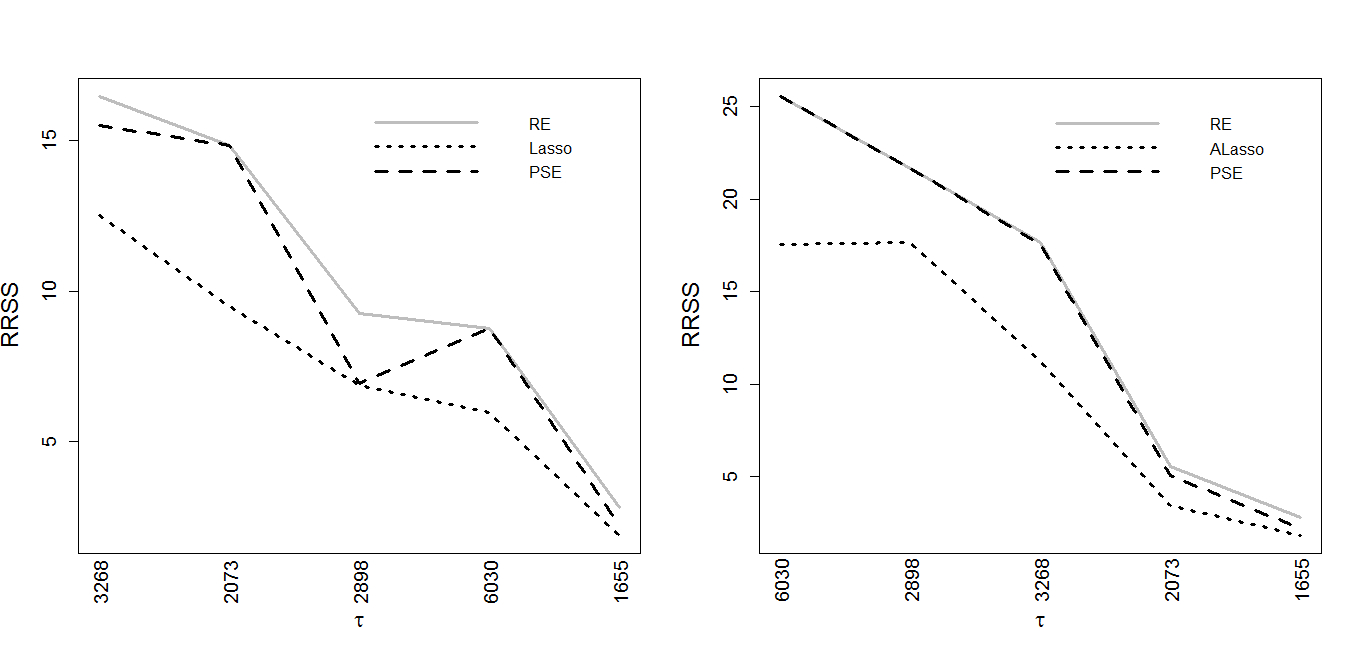}}
 %     \scalebox{0.4}[0.4]{\includegraphics{econ_pmse_alasso.PNG}}
 $$
  \caption{RRSS from \eqref{eq-mser} from post selection PSE and
  the Lasso-type estimators: Lasso (left panel)
   or Adaptive Lasso (ALasso) (right panel).
   The curve is plotted based upon a decreasing order of RRSS
   for better visibility, with corresponding values of $\tau$ plotted in
     X-axis.
}
 \label{fig-bl-mser1}
\end{figure}

Since we do not know what the true model is in the real data analysis,
we first evaluate the prediction improvement
from variable selection estimates to  post selection PSEs by
 computing the relative residual sum of squares (RRSS) of
 the estimator $\bbeta_\cJ^*$ over the weighted ridge estimator $\hbbetaWR_\cJ$ as follows
\bel{eq-mser}
{\rm  RRSS}(\bbeta^*_{\cJ})=  \dfrac{\sum_{i=1}^n \|\by-\sum_{j\in \cJ} \bX_{\cJ}\hbbetaWR_{\cJ}\|^2}
{\sum_{i=1}^n \|\by-\sum_{j\in \cJ} \bX_{\cJ}\bbeta^\diamond_{\cJ}\|^2 },
\eel
where $\cJ$ is the index of the submodel chosen
by corresponding variable selection methods, and
$\bbeta^\diamond_{\cJ}$ can be (Adaptive) Lasso
and the corresponding generated post selection SEs and post selection PSEs.
Similar to the simulation studies, ${\rm RRSS}>1$ indicates the superiority of
$\bbeta^*_{\cJ}$ over $\hbbeta_{\cJ}$.
The results on RRSS for different $\tau$'s
are reported in Figure \ref{fig-bl-mser1},
where the left and right panels are based upon Lasso and
Adaptive Lasso submodels, respectively.
Those RRSS  values of post selection REs give the highest value in both cases. This is not surprising since
we assume  the selected submodel is the right one
and does not account for any bias.
In both cases, the post selection PSEs dominates the
  corresponding variable selection estimation
  in terms of the RRSS no matter whether Lasso or
Adaptive Lasso is used for generating the candidate submodel.
This is because shrinkage estimation
provides a better trade-off between bias and variance
when selected submodels underfit the true model.

\begin{figure}[tp]
\centering
 %$$
%   \scalebox{0.4}[0.4]{\includegraphics{econ_boxplot_lasso.png}}
% $$
 $$
   \scalebox{0.4}[0.4]{\includegraphics{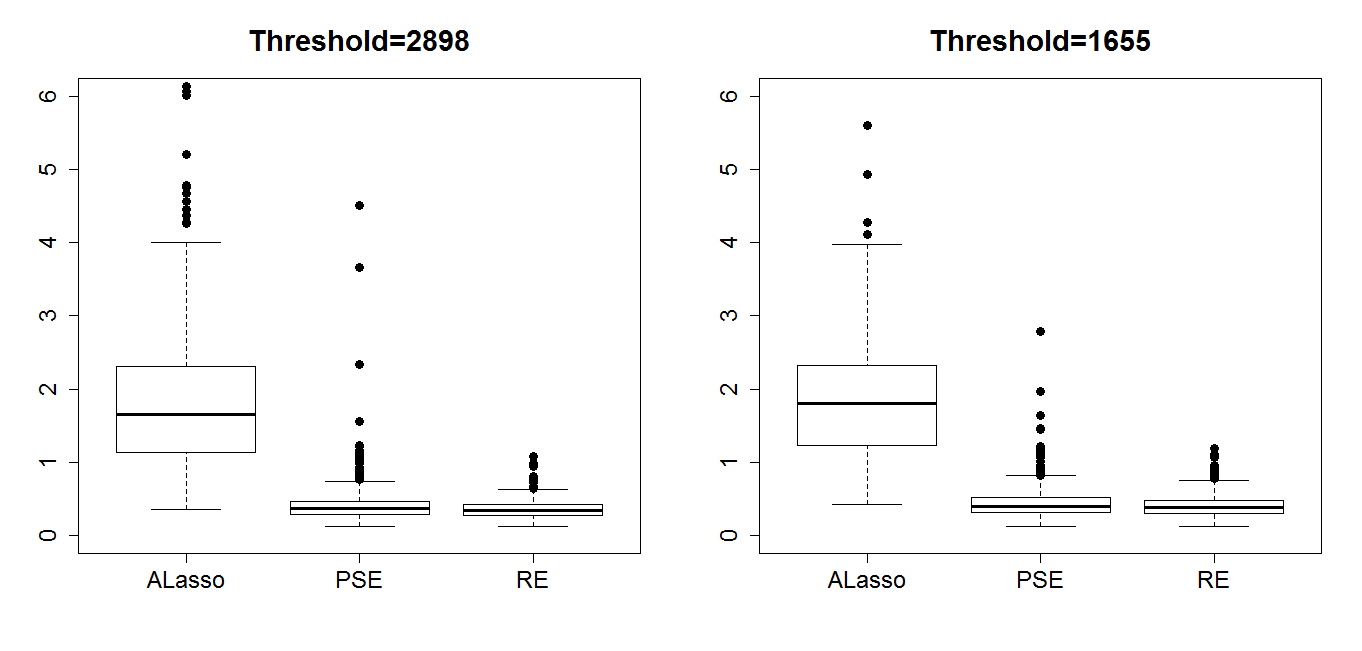}}
  $$
  \caption{Prediction errors from post selection PSE, RE and Adaptive Lasso (ALasso).
 Left: $\tau=2898$; Right: $\tau=1655$.
       All  prediction errors are  computed using cross validation following 500 random partitions
    of the data set. In each partition, the training set consists of 2/3
observations and the test set consists of the remaining 1/3 observations.
 }
 \label{fig-bl-mser2}
\end{figure}

In addition, we also obtain prediction errors
using cross validation following 500 random partitions
    of the data set. In each partition, the training set consists of 2/3
observations (size 55) and the test set consists of the remaining 1/3 observations (size 28).
Corresponding results for  $\tau=2898$ and $1655$  are reported in  Figure
 \ref{fig-bl-mser2}, where the post selection PSEs is compared with the Adaptive Lasso.
 The comparisons between  the post selection PSEs and (Adaptive) Lasso
for other $\tau$'s  follow the similar pattern, and thus are omitted.
It is observed that   post selection PSEs  produce  much smaller prediction errors than
the Lasso-type estimation.

\begin{table} [h]
 \begin{center}
   \caption {List of Variable}
    \label{data-table1}
  {\small  \begin{tabular} { ll   } \\ \hline\hline
  Variable & Description\\ \hline
  Dependent Variable &\\
gr & Annualized GDP growth rate in the period of 1960--85\\
Threshold Variables &\\
gdp60 & Real GDP per capita in 1960 (1985 price)\\
%lr & Adult literacy rate in 1960\\
Covariates&\\
lgdp60&  $\log$ GDP per capita in 1960 (1985 price)\\
%lr &Adult literacy rate in 1960 (only included when Q = lr)\\
lsk &Log(Investment/Output) annualized over 1960-85; \\
    &a proxy for the log physical  savings rate\\
lgrpop  &  $\log$ population growth rate annualized over 1960--85\\
pyrm60  &  $\log$ average years of primary schooling in the male population in 1960\\
pyrf60  &  $\log$ average years of primary schooling in the female population in 1960\\
syrm60  &  $\log$ average years of secondary schooling in the male population in 1960\\
syrf60  &  $\log$ average years of secondary schooling in the female population in 1960\\
hyrm60  &  $\log$ average years of higher schooling in the male population in 1960\\
hyrf60  &  $\log$ average years of higher schooling in the female population in 1960\\
nom60  & Percentage of no schooling in the male population in 1960\\
nof60  & Percentage of no schooling in the female population in 1960\\
prim60  & Percentage of primary schooling attained in the male population in 1960\\
prif60  & Percentage of primary schooling attained in the female population in 1960\\
pricm60  & Percentage of primary schooling complete in the male population in 1960\\
pricf60  & Percentage of primary schooling complete in the female population in 1960\\
secm60  & Percentage of secondary schooling attained in the male population in 1960\\
secf60  & Percentage of secondary schooling attained in the female population in 1960\\
seccm60  & Percentage of secondary schooling complete in the male population in 1960\\
seccf60  & Percentage of secondary schooling complete in the female population in 1960\\
llife  &  $\log$ of life expectancy at age 0 averaged over 1960--1985\\
lfert  &  $\log$ of fertility rate (children per woman) averaged over 1960--1985\\
edu/gdp  & Government expenditure on eduction per GDP averaged over 1960--85\\
gcon/gdp  &  Government consumption expenditure net of defence and education\\
        & per GDP averaged over 1960--85\\
revol  & The number of revolutions per year over 1960--84\\
revcoup &  The number of revolutions and coups per year over 1960--84\\
wardum  &  Dummy for countries that participated in at least one external war over
1960--84\\
wartime  &  The fraction of time over 1960-85 involved in external war\\
lbmp  & $\log$(1+black market premium averaged over 1960--85)\\
tot  & The term of trade shock\\
lgdp60 & ``educ'' Product of two covariates (interaction of lgdp60 and education\\
& variables from pyrm60 to seccf60); total 16 variables\\
\\
\hline \hline
   \end{tabular}}
  \end{center}
  \end{table}

\begin{table} [h]
 \begin{center}
   \caption {Sizes of  Selected Submodel}
    \label{data-table2}
  {\small  \begin{tabular} {  c  c c c  c c  } \\ \hline\hline
  $\tau$     &$6030 $&$ 3268 $&$ 2898 $&$ 2073 $&$ 1655$  \\ \hline
  Lasso &15 & 18 & 18 & 19 & 11  \\
  Adaptive  Lasso    &19& 13& 20& 19& 11\\
  % 19 13 20 19 11
\hline \hline
   \end{tabular}}
  \end{center}
  \end{table}

\begin{table} [h]
 \begin{center}
   \caption { Frequency of selected variables (based upon either $\beta_j\neq 0$ or $\delta_j\neq 0$)
   among All 5 $\tau$'s}
    \label{data-table3}
  {\small  \begin{tabular} {  c   cccc }
   \\ \hline\hline

   &\multicolumn{2}{c}{ Lasso}&\multicolumn{2}{c}{ ALasso} \\
    \cline{2-3} \cline{4-5}\\
  % &Frequency  &Frequency  \\ \hline
  {Variable}           &  \#($\beta_j\neq 0$) & \#($\delta_j\neq 0$)
              &  \#($\beta_j\neq 0$) & \#($\delta_j\neq 0$)\\ \hline
lgdp60       & 5  &  0&  5  & 0\\
lsk         &  5  &  0 & 5  &  0\\
nom60       &   0  &  1&   0  & 1\\
prim60       &  3  &  0&  3 &  0\\
pricm60      & 3   & 3&  3   & 3\\
seccm60    &    0  &  5&   0  &  5\\
seccf60    &    1  &  0&    1  &  0\\
llife       &  5  &  0&   5  &  0\\
lfert      &   5   & 0&   5  &  0\\
edugdp       & 3   & 0&   4  &  0\\
gcongdp      & 5  &  0& 5  & 0\\
revol        &  2 &   0&  3 &   0\\
wardum       &  2  & 3 &  2   & 3\\
wartime      & 4  &  4&  3  & 3\\
lbmp         &  5  &  0&  5  &  0\\
tot          & 0 &  5&  0  &  5\\
lgdpsyrm60   & 2  & 0& 2  & 0\\
lgdphyrm60   &  3   & 0& 1   & 0\\
lgdphyrf60   & 0  &  1&  1  &  0\\
lgdpnof60    &  0  & 3&   0  & 3\\
lgdpprim60    & 2   & 0& 2   & 1\\
lgdpprif60    & 0  &  1& 0  & 2\\
lgdpseccf60  &  1  &  0& 0 &  0\\
     \hline      \hline
   \end{tabular}}
  \end{center}
  \end{table}

\begin{table} [h]
 \begin{center}
   \caption {Estimation results under  $\tau=2898$ (Candidate submodel from ALasso)}
    \label{data-table4}
  {\small  \begin{tabular} {  lcccc }
   \\ \hline\hline
  {Variable}  &$\hbeta^{(ALasso)}$ & $\widehat{\delta}^{(ALasso)}$      &$\hbeta^{(PSE)}$ & $\widehat{\delta}^{(PSE)}$\\
\hline
lgdp60   &   $-9.253\times 10^{-3}$ &$-$ & $-1.287\times 10^{-2}$  &$-$ \\
lsk       &   $6.121\times 10^{-4}$  &$-$ &  $3.942\times 10^{-4}$   &$-$ \\
nom60        &$-$ &  $1.400\times 10^{-2}$  &$-$ &  $3.481\times 10^{-2}$\\
prim60     & $-4.579\times 10^{-2}$  &$-$ & $-7.472\times 10^{-2}$  &$-$ \\
pricm60   &  $1.934\times 10^{-2}$ & $ 1.974\times 10^{-3}$ & $4.129\times 10^{-2}$ & $7.058\times 10^{-3}$\\
seccm60      &$-$ &  $4.903\times 10^{-4}$   &$-$ & $ 4.324\times 10^{-4}$ \\
llife       & $1.200\times 10^{-3}$  &$-$ & $ 2.212\times 10^{-3}$  &$-$ \\
lfert       & $-1.659\times 10^{-3}$  &$-$ & $-1.507\times 10^{-3}$  &$-$ \\
edugdp      & $ 2.228\times 10^{-5}$  &$-$ & $ 2.309\times 10^{-5}$  &$-$ \\
gcongdp    & $ -2.351\times 10^{-4}$   &$-$ & $-2.610\times 10^{-4}$   &$-$ \\
revol      & $ -1.020\times 10^{-6}$  &$-$ & $-1.158\times 10^{-4}$   &$-$ \\
wardum       &$-$ & $-1.417\times 10^{-4}$  &$-$ & $-3.336\times 10^{-4}$\\
wartime    & $ -1.655\times 10^{-4}$   &$-$ & $-5.081\times 10^{-5}$ &$-$ \\
lbmp       & $ -1.580\times 10^{-3} $ &$-$ & $-1.595\times 10^{-3}$ &$-$ \\
tot          &$-$ &  $5.202\times 10^{-6}$  &$-$ &  $6.318\times 10^{-6}$\\
lgdphyrm60  & $ 1.122\times 10^{-2}$  &$-$ & $ 4.291\times 10^{-2}$ &$-$ \\
lgdphyrf60  & $-7.585\times 10^{-3}$  &$-$ & $-4.143\times 10^{-2}$  &$-$ \\
lgdpnof60    &$-$ & $ 6.392\times 10^{-2}$  &$-$ & $ 0.189$\\
lgdpprif60   &$-$ & $-3.130\times 10^{-2}$  &$-$ & $-0.127$\\

     \hline      \hline
   \end{tabular}}
  \end{center}
  \end{table}

\begin{table} [h]
 \begin{center}
   \caption {Estimation results under $\tau=1655$ (Candidate submodel from ALasso)}
    \label{data-table5}
  {\small  \begin{tabular} {  lcccc }
   \\ \hline\hline
  {Variable}  &$\hbeta^{(ALasso)}$ & $\widehat{\delta}^{(ALasso)}$      &$\hbeta^{(PSE)}$ & $\widehat{\delta}^{(PSE)}$\\
\hline
lgdp60     & $ -2.841\times 10^{-3}$& $  -$ & $-1.306\times 10^{-2}$& $  -$\\
lsk        & $  1.319\times 10^{-3}$ & $ -$ & $ 1.284\times 10^{-3}$ & $ -$\\
seccm60    & $  -$& $  3.652\times 10^{-4}$ & $ -$ & $ 5.873\times 10^{-4}$\\
llife      & $  3.532\times 10^{-4}$& $  -$ & $ 1.633\times 10^{-3}$ & $ -$\\
lfert      & $ -2.552\times 10^{-4}$ & $ -$ & $-2.250\times 10^{-3}$  & $-$\\
gcongdp    & $ -1.554\times 10^{-4}$ & $ -$& $ -3.033\times 10^{-4}$ & $ -$\\
revol      & $ -3.715\times 10^{-5}$& $ -$ & $-9.248\times 10^{-4}$ & $-$\\
wartime    & $ -4.965\times 10^{-5}$& $ -1.120\times 10^{-5}$& $  2.731\times 10^{-4}$& $ -3.958\times 10^{-5}$\\
lbmp      & $  -1.428\times 10^{-3}$ & $ -$ & $-5.887\times 10^{-4}$ & $ -$\\
tot        & $  -$ & $ 5.175\times 10^{-7}$ & $ -$& $  8.476\times 10^{-6}$\\
     \hline      \hline
   \end{tabular}}
  \end{center}
  \end{table}

  \section{Conclusion and discussions}
In this paper, we generalize the  shrinkage
estimation to a high-dimensional sparse regression model.
We propose a  post selection  shrinkage estimation strategy
 by shrinking  a weighted
ridge  estimator in the direction of a candidate submodel
obtained by existing penalized least squares variable selection methods.

When $p_n$ grows with $n$ quickly, it is reasonable  to
assume that  the model
sparsity exists in the sense that most  covariates do not contribute. However, at the same time,
some covariates may still make some small but jointly non-trivial  contribution to the
response. Existing penalized regularization approaches
usually lead to a sparse model,
but tends to miss the possible small contributions from some covariates,
resulting in excessive prediction errors or inefficient estimation.
Our proposed post selection shrinkage strategy,
taking into account possible contributions of
   covariates with weak and/or moderate signals,
  has dominant prediction
performances over candidate submodel estimates generated from
Lasso-type methods.

Before obtaining a shrinkage estimator, one key step is
to generate a full estimation of $\bbeta_n$ when $p\gg n$.
We suggest a post selection weighted ridge estimator which is able to
separate  small coefficients from zero coefficients.
 %when prior information is available.
 The advantages of proposed post selection PSE
 are studied both theoretically and numerically.
In theory, we established
the asymptotic normality of
the post selection weighted ridge estimator when
$p_n$ grows with $n$ at an almost exponential rate such that $\log(p_n)=O(n^\nu)$ for some $0<\nu<1$. Those novel
asymptotic properties are used for investigating
the asymptotic efficiency of the proposed post selection PSE analytically.
 In numerical studies, we chose tuning parameters $c_1$ and $c_2$
from cross validation , but cannot guarantee their optimality for post selection PSE.
 The choice of tuning parameters
 is an important but challenging issue in HD data analysis
 which could potentially create very important future work.
Although the proposed post selection PSE
 was presented based on  a weighted ridge method, other methods can also be used to generate the shrinkage estimator.
%
%  any other full estimation enjoying good  asymptotic normality properties under
%some regular conditions can be used to generate
%a shrinkage estimator.
%In computation, we suggest the use of a weighted ridge
%estimator to generate a full estimation because of the
%computation's convenience.
%In a possibly ultrahigh dimension ($p \gg n$) estimation problem, we make the assumption
%that a successful reduction to a problem with much lower dimensions ($p = O(n^\nu)$) exists
%with some $\nu\ge 1$.

Finally, we acknowledge the importance of Lasso-type variable selection methods, but at the same time,
do not depend completely on  them, especially when many weak coefficients jointly affect the response variable.
The Lasso is the start, but not the end.
We could potentially still make some significant prediction improvements. We hope this work will
 shed some more light on the investigation of the post variable selection shrinkage analysis in high-dimensional data analysis.

%\begin{appendix}
\section*{Appendix}
All technical proofs are given in this section.

\noindent{\bf Proof of Theorem  1}

After solving \eqref{part-ridge-obj},  we obtain
\bel{beta-uer1}
\tbbeta_{\widehat S_1}(r_n)=(\bX_{\widehat S_1}'\bM_{\widehat S_1^c}(r_n)\bX_{\widehat S_1})^{-1} \bX'_{\widehat S_1} \bM_{\widehat S_1^c}(r_n)\by
\eel
 and
 \bel{beta-uer2}
\tbbeta_{\widehat S_1^c}(r_n)=(r_n\bI_{p_{2n}}+\bX_{\widehat S_1^c}'\bM_{\widehat S_1}\bX_{\widehat S_1^c})^{-1} \bX'_{\widehat S_1^c} \bM_{\widehat S_1} \by,
\eel
where  $\bM_{\widehat S_1^c}(r_n)=\bI_n-\bX_{\widehat S_1^c}(r_n\bI_{p_{2n}}+\bX_{\widehat S_1^c}'\bX_{ \widehat S_1^c})^{-1}\bX_{\widehat S_1^c}'$
and $\bM_{\widehat S_1}=\bI_n-\bX_{\widehat S_1}(\bX_{\widehat S_1}'\bX_{\widehat S_1})^{-1}\bX_{\widehat S_1}'$.

We only need to prove the result under the condition $\widehat S_1=S_1$. Then  all matrices, vectors indexed
by $\widehat S_1$ can be replaced by $S_1$ or $1$ without causing of any confusion.
For example, $\bM_{\widehat S_1}=\bM_{S_1}=\bM_{1}$ under the condition.

First, we check the bias of $\hbbetaWR_{S_1^c}$.
Since $\bM_1$ is an idempotent matrix, $\bM_1\bX_{1n}=0$. Denote $q_n=p_{2n}+p_{3n}$. Then
$$(\bX_{S_1^c}'\bM_1\bX_{S_1^c}+r_n\bI_{q_n})^{-1}\bX_{S_1^c}'\bM_1\bX_{1n}\bbeta_{10}=\bzero.$$
Let $\bQ$ be a $q_n\times q_n$ orthogonal matrix  such that
$$\bU'\bM_1\bU=\bX_{S_1^c}'\bM_1\bX_{S_1^c}=\bQ
 \begin{pmatrix}
 \bD &\bzero\\
 \bzero &\bzero
 \end{pmatrix} \bQ', $$
where $\bD=\diag\{\varrho_{1n},\cdots,\varrho_{k_n n}\}$. Then we have
\bel{eq-thm1-bias}
\begin{array}{ll}
E(\hbbetaWR_{S_1^c})-\bbeta^*_{S_1^c}&=(\bX_{S_1^c}'\bM_1\bX_{S_1^c}+r_n\bI_{q_n})^{-1}\bX_{S_1^c}'\bM_1\by-\bbeta^*_{ S_1^c}]\\
&=(\bX_{S_1^c}'\bM_1\bX_{S_1^c}+r_n\bI_{q_n})^{-1}\bX_{S_1^c}'\bM_1\bX_{S_1^c}\bbeta^*_{ S_1^c}-\bbeta^*_{ S_1^c}\\
&=-r_n(\bX_{S_1^c}'\bM_1\bX_{S_1^c}+r_n\bI_{q_n})^{-1}\bbeta^*_{ S_1^c}\\
&=-\bQ  \begin{pmatrix}
 (\bI_{k_n}+r_n^{-1}\bD)^{-1} &\bzero\\
 \bzero &\bI_{q_n-k_n}
 \end{pmatrix} \bQ'\bbeta^*_{ S_1^c}.
\end{array}
\eel
 Suppose that $\bQ=(\bQ_1, \bQ_2)$ and $\bQ_1$ is a $q_n\times{k_n}$ matrix.
Notice that $\bQ\bQ'=\bQ'\bQ=\bI_{q_n}$.
Then $\bQ_1'\bQ_1=\bI_{k_n}$, $\bQ_1'\bQ_2=\bzero$,  and  $\bQ_2\bQ_2'$ is a projection matrix.
Let  $\btheta^*=\bQ_1\bQ_1'\bbeta^*_{ S_1^c}$. Then
\bel{eq-theta-beta}
\bbeta^*_{ S_1^c}=\bQ_1\bQ_1'\btheta^*=\bQ_1\bQ_1'\bQ_1\bQ_1'\bbeta^*_{ S_1^c}=\bQ_1\bQ_1'\bbeta^*_{ S_1^c}.
\eel
Replace $\bbeta^*_{ S_1^c}$ in \eqref{eq-thm1-bias} by $\bQ_1\bQ_1'\bbeta^*_{ S_1^c}$, we have
$$
%\begin{array}{ll}
E(\hbbetaWR_{S_1^c})-\bbeta_{0 S_1^c}= -\bQ_1 (\bI_{k_n}+r_n^{-1}\bD)^{-1}\bQ_1'\bbeta^*_{ S_1^c}.
%\end{array}
$$
Thus,
$$
\|E(\hbbetaWR_{S_1^c})-\bbeta*_{S_1^c}\|^2=\btheta_0'\bQ_1(\bI_{k_n}+r_n^{-1}\bD)^{-2}\bQ_1\btheta_0
\le (1+\varrho_{1n}/r_n)^{-2}\|\bbeta_{0 S_1^c}\|^2
$$
For every $j\notin S_{1}$, $|{\rm bias}(\hbetaWR_j)|\le \|E(\hbbetaWR_{S_1^c})-\bbeta_{0 S_1^c})\|$,
and thus
$$|{\rm bias}(\hbetaWR_j)| \le(1+\varrho_{1n}/r_n)^{-1}\|\bbeta_{0 S_1^c}\|
   \le (r_n/\varrho_{1n})O(n^\tau)\le O(r_n n^{\tau-\eta}).$$
   The rest of the proof just mimics the proof of Theorem 2 in \cite{shao.deng:2012}.
   We will provide some outlines of the the proof.
If we let $r_n=c_2 a_n^{-2}(\log\log n)^3\log(n\vee p)$ and $\log(p_n)=O(n^\nu)$ in (B3),
then for $u_n=1+(\log\log n)^{-1}$, we have
$$
\dfrac{|{\rm bias}(\hbetaWR_j)|}{a_n(u_n-1)}
\le \dfrac{r_n n^{\tau-\eta}}{a_n(u_n-1)}
\le \dfrac{c_2(\log\log n)^4}{a_n^3 n^{\eta-\tau-\nu}}
\le  \dfrac{c_2(\log\log n)^4}{c_1^3 n^{\eta-\tau-\nu-3\alpha}}
\to 0 \quad{\rm if~} 3\alpha<\eta-\nu-\tau,
$$
where the last ``$\le$'' is from \eqref{eq-an} and $c_1$ is defined there.
From the normal assumption of $\veps_i$ and
the solution in linear  expression in \eqref{beta-uer2}, we know
$\hbbetaWR_{S_1^c}$ is normally distributed and
 $$
 \begin{array}{ll}
 {\rm Var}(\hbbetaWR_{S_1^c})&=\sigma^2 (\bX_{S_1^c}'\bM_1\bX_{S_1^c}+r_n\bI_{q_n})^{-1}\bX_{S_1^c}'\bM_1\bX_{S_1^c}(\bX_{S_1^c}'\bM_1\bX_{S_1^c}+r_n\bI_{q_n})^{-1}\\
 &\preceq \sigma^2 (\bX_{S_1^c}'\bM_1\bX_{S_1^c}+r_n\bI_{q_n})^{-1}\\
 & \preceq  \sigma^2 r_n^{-1} \bI_{q_n},
   \end{array}
  $$
  where ``$\bA\preceq\bB$'' means $\bB-\bA$ is a non-negative definite matrix.
 Thus for any $j\notin S_1$, ${\rm Var}(\hbetaWR_j)= O(1/r_n)$.
 Notice that $\sqrt{r_n}a_n(u_n-1)=O\left((\log\log n)^{1/2}\right)\to \infty$.
We have
$$
\dfrac{a_n(u_n-1)}{\sqrt{{\rm Var}(\hbetaWR_j)}}\ge a_n(u_n-1)\sqrt{r_n}\to \infty.
$$

 $$
 \begin{array}{ll}
% P\left(\dfrac{}{\sqrt{\var(\hbetaWR_j)}}\right)
P\left(|\hbetaWR_j-\beta^*_j|>a_n(u_n-1)\right)&\le
P\left(|N(0,1)|>\dfrac{a_n(u_n-1)}{\sqrt{{\rm Var}(\hbetaWR_j)}}-\dfrac{|{\rm bias}(\hbetaWR_j)|}{\sqrt{{\rm Var}(\hbetaWR_j)}}\right)\\
&=  2 \Phi \left(\dfrac{|{\rm bias}(\hbetaWR_j)|-a_n(u_n-1)}{\sqrt{{\rm Var}(\hbetaWR_j)}} \right)\\
&\le  2 \Phi \left( - c_0 \sqrt{r_n} a_n/(\log\log n)\right)\\
&\le \exp\{-c_0^2 r_n a_n^2 / (\log\log n)^2\},
 \end{array}
 $$
 where $\Phi$ is the cumulative distribution function of a standard normal random variable,
 $c_0>0$ is a constant, ``$\le$'' is the tail probability of a normal random variable.
 Thus,
  $$
 \begin{array}{ll}
 &P\left(\{j\notin S_1: |\beta^*_j|>a_n u_n\}\subset \{j\notin S_1: |\hbetaWR_j|>a_n\} \right)\\
 &\ge 1-P\left(\displaystyle\bigcup_{j: |\beta^*_j|>a_n u_n}  \{|\hbetaWR_j|\le a_n\} \right)\\
  &\ge 1-P\left(\displaystyle\bigcup_{j: |\beta^*_j|>a_n u_n}  \{|\hbetaWR_j-\beta_j^*|\le a_n(u_n-1)\} \right)\\
 &\ge 1-\displaystyle\sum_{j\notin S_1} P\left(|\hbetaWR_j-\beta^*_j|>a_n(u_n-1)\right)\\
 &\ge 1-q_n \exp\{-c_0^2 r_n a_n^2 / (\log\log n)^2\}\\
 &\ge 1- \exp\{-\left(c_0^2 r_n a_n^2 / (\log\log n)^2-\log(p_n)\right)\}\\
 &\ge 1-\exp\{- (c_0^2 \log\log n -1)\log(p_n\vee n)\}.
  \end{array}
 $$
When $n$ is large enough, there exists $c_0^2 \log\log n -1> t>0$ for some $t>0$.
 Thus $$\lim_{n\to \infty} P\left(\{j\notin S_1: |\beta^*_j|>a_n u_n\}\subset \{j\notin S_1: |\hbetaWR_j|>a_n\} \right)
      \ge 1-(p_n\vee n)^{-t}\to 1.$$
Similarly we have
 $$\lim_{n\to \infty} P\left(\{j\notin S_1: |\beta^*_j|>a_n/u_n\}\supset \{j\notin S_1: |\hbetaWR_j|>a_n\} \right)\ge 1-(p_n\vee n)^{-t}\to 1.$$
Because of the continuity of $\hbetaWR_j$ and $\lim_{n\to \infty}u_n= 1$,
 we have
 $$
 \lim_{n\to \infty} P(\widehat S_2|\widehat S_1=S_1)=1.
 $$
$\Box$

\vspace{10pt}
\noindent{\bf Proof of Corollary \ref{cor2}}

Since $S_1\subset \widehat S_1$,
 a weighted ridge estimator $\hbbeta_{\widehat S_1}$ aims to find some weak signals from  $\widehat S_1^c\cap S_2$.
 Since $\widehat S_1^c\subset S_1^c$,
 the smallest positive eigenvalues of $\bX_{\widehat S_1^c}'\bM_{\widehat S_1}\bX_{\widehat S_1}$
must be larger than $\lm_{1n}$, and
$\|\bbeta^*_{\widehat S_1^c}\|_2\le \|\bbeta^*_{S_1^c}\|_2$. Thus we can
borrow the proof of Theorem 1 here, by treating $\widehat S_1$ and $S_2\cap \widehat S_1^c$
as the new $S_1$ and $S_2$.
$\Box$

%To prove Theorem 2,
%we first notice that $\hbbetaWR_{S_1^c}$ is consistent to any linear combination of $\bbeta_{(-1)0}$
%as given in the following lemma.
%The proof is omitted since it is similar to one for Theorem 1 in Shao and Deng (2012).

%\begin{lemma}\label{Lemma 2}
%Under conditions in Theorem  1, for any vector $\bd$ such that $\|\bd\|=1$,
%we uniformly have
%$$
%E(\bd'\hbbetaWR_{S_1^c}-\bd'\bbeta_{0 S_1^c})^2 =O(r_n^{-1})+O(r_n^2 n^{-2(\eta-\tau)})
%$$
%for $r_n\ge O(n^{2(\eta-\tau)/3})$.
%\end{lemma}

\vspace{10pt}
\noindent{\bf Proof of Theorem 2}

Similar to the proof in Theorem 1, we start the proof by assuming $\widehat S_1=S_1$.
Then the penalized quadratic loss function in \eqref{part-ridge-obj} becomes
$$
L(\bbeta_n; S_1)=\{\|\by-\bX_n\bbeta\|^2 +r_n\|\bbeta_{S_1^c}\|^2  \}.
$$
Therefore, $\hbbetaWR_n=\argmin\{L(\bbeta_n; S_1)\}$ satisfies,
$$
\frac{\partial L(\hbbetaWR_n)}{\partial \bbeta_{S_3^c}}=\bzero.
$$
From the notation $\bX_{S_3^c}'=\bZ=(\bz_1,\cdots,\bz_n)$. If we
write  $\bX_{S_3}'=(\bw_1,\cdots,\bw_n)$, then
$$
-\sum_{i=1}^n(y_i-\bz_i'\hbbetaWR_{S_3^c}- \bw_i' \hbbetaWR_{S_3})\bz_i +r_n
\begin{pmatrix}
\bzero_{p_{1n}}&\\
 \hbbetaWR_{S_2}
\end{pmatrix}
=\bzero_{p_{1n}+p_{2n}}.
$$
Replacing $y_i$ by $\bz_i'\bbeta_{0 S_3^c}+\bw_i'\bbeta_{0 S_3}+\veps_i$, we have
$$
-\sum_{i=1}^n(\veps_i-\bz_i'(\hbbetaWR_{ S_3^c}-\bbeta_{0 S_3^c})- \bw_i' (\hbbetaWR_{S_3}-\bbeta_{0S_3}))\bz_i +r_n
\begin{pmatrix}
\bzero_{p_{1n}}&\\
 \hbbetaWR_{S_2}
\end{pmatrix}
=\bzero.
$$
Notice  that $\bSigma_{n} =n^{-1}\sum_{i=1} \bz_i\bz_i'\to \bSigma$. Thus,
\bel{eq-betahat1}
\begin{array}{ll}
n^{1/2}\bd_n'(\hbbetaWR_{S_3^c}-\bbeta^*_{S_3^c})&= n^{-1/2} \sum_{i=1}^n \bd_n' \veps_i\bSigma_{n}^{-1}\bz_i -n^{-1/2} r_n \bd_n'\bSigma_{n}^{-1}
\begin{pmatrix}
\bzero_{p_{1n}}&\\
 \hbbetaWR_{2n}
\end{pmatrix}\\
 &\quad - n^{-1/2}\sum_{i=1}^n \bd_n' \bw_i' (\hbbetaWR_{3n}-\bbeta^*_{S_3}) \bSigma_{n}^{-1} \bz_i\\
 \end{array}
\eel
Under conditions (B1-B3), with probability 1,  $\hbbetaWR_{S_3}=\bzero$ from Theorem 1. Therefore, the third term
 in \eqref{eq-betahat1}  is zero.
 By abusing the notation, if we rewrite $\bd_{n}=(\bd_{1n}',\bd_{2n}')'$,
 then
 $$
 \begin{array}{ll}
 n^{-1/2} r_n \bd_n'\bSigma_{n}^{-1}
 \begin{pmatrix}
\bzero_{p_{1n}}&\\
 \hbbetaWR_{S_2}
\end{pmatrix}
&\le \rho_{1}^{-1} n^{-1/2} r_n \bd_{2n}'\hbbetaWR_{S_2} \\
& =O_P(\rho_{1}^{-1} n^{-1/2} r_n \bd_{2n}'\bbeta^*_{S_2})\\
& \le O_P\left(\rho_{1}^{-1}r_n n^{-1/2} \|\bd_{2n}\|\|\bbeta_{S_2}^*\|\right)\\
&\le O_P\left(\rho_{1}^{-1}r_n n^{-(1/2-\tau)}\right)=o_P(1),
\end{array}
 $$
where the first ``$\le$'' is from (B4), the
first ``='' is from \eqref{beta-uer2} and (B1),
the second ``$\le$'' is from Cauchy-Schwarz inequality,
the third ``$\le$'' is from   (A2). The
last ``='' holds since $r_n=o(n^{1/2-\tau})$ if
 we choose $r_n=c_2 a_n^{-2}(\log\log n)^3 \log(n\vee p_{n})$ with
 $a_n=c_1 n^{-\alpha}$ for  $\alpha<1/4-\tau/2$ for $0<\tau<1/2$.
Therefore,
\bel{eq-betahat2}
\lim_{n\to \infty} n^{1/2}s_n^{-1}\bd_n'(\hbbetaWR_{S_3^c}-\bbeta^*_{ S_3^c})
= \lim_{n\to \infty} n^{-1/2} s_n^{-1}\sum_{i=1}^n \bd_n' \veps_i\bSigma_{n}^{-1}\bz_i
\eel
Define  $u_{i}=n^{-1/2}s_n^{-1}\bd_n' \bSigma_{n}^{-1}\bz_i , ~1\le i\le n$.
From (B1), we know that $\sum_{i=1}^n u_i \veps_i$ is  normal
with variance,
$$
\Var\left( \sum_{i=1}^n (u_{i}\veps_i)\right)
=\sigma^2   n^{-1}s_n^{-2}\bd_n' \bSigma_{n}^{-1}(\sum_{i=1}^n\bz_i\bz_i) \bSigma_{n}^{-1}\bd_n
=1.
$$
$\Box$

\vspace{10pt}
\noindent{\bf Proof of Theorem 3}

First \eqref{aqr-ue} holds since we have
$$
\lim_{n\to\infty} E[n^{1/2}s_{1n}^{-1}\bd_{1n}'(\hbbetaWR_{1n}-\bbeta^*_{1})]^2]=
E\{\lim_{n\to\infty} [n^{1/2}s_{1n}^{-1}\bd_{1n}'(\hbbetaWR_{1n}-\bbeta^*_{1})]^2]\}=E[Z^2]=1,
$$
where $Z\sim N(0,1)$.
We now verify \eqref{aqr-re}.
Let $\widetilde\by=\by-\bX_{2n}\hbbetaWR_{2n}-\bX_{3n}\hbbetaWR_{3n}.$
Then
\bel{beta1-new-residual}
\begin{array}{ll}
\hbbetaWR_{1n}&=\argmin\{\|\widetilde\by-\bX_{1n}\bbeta_{1n}\|^2\}\\
&=(\bX_{1n}'\bX_{1n})^{-1}\bX_{1n}'\widetilde\by\\
%&=(\bX_{1n}'\bX_{1n})^{-1}\bX_{1n}\by-(\bX_{1n}'\bX_{1n})^{-1}\bX_{1n}'\bX_{2n}\hbbetaWR_{2n}-(\bX_{1n}'\bX_{1n})^{-1}\bX_{1n}'\bX_{2n}\hbbetaWR_{3n}\\
&=\hbbetaRE_{1n}-(\bX_{1n}'\bX_{1n})^{-1}\bX_{1n}'\bX_{2n}\hbbetaWR_{2n}-(\bX_{1n}'\bX_{1n})^{-1}\bX_{1n}'\bX_{3n}\hbbetaWR_{3n}\\
&=\hbbetaRE_{1n}-(\bX_{1n}'\bX_{1n})^{-1}\bX_{1n}'\bX_{2n}\hbbetaWR_{2n}.
\end{array}
\eel
From the definition,
$$
\begin{array}{ll}
  R(\bd_{1n}'\hbbetaRE_{1n})&=\lim_{n\to\infty} E[n^{1/2}s_{1n}^{-1}\bd_{1n}'(\hbbetaRE_{1n}-\bbeta^*_{1}) ]^2\\
 &=  \lim_{n\to\infty}s_{1n}^{-2} E\{n^{1/2}\bd_{1n}'[(\hbbetaWR_{1n}-\bbeta^*_{1})-(\hbbetaWR_{1n}-\hbbetaRE_{1n})]\}^2\\
 &=\lim_{n\to\infty} E\{n^{1/2}s_{1n}^{-2}\bd_{1n}'(\hbbetaWR_{1n}-\bbeta^*_{1})\}^2
   +\lim_{n\to\infty} E\{n^{1/2}s_{1n}^{-2}\bd_{1n}'(\hbbetaWR_{1n}-\hbbetaRE_{1n})\}^2\\
  &\quad -2 \lim_{n\to\infty} E \{ns_{1n}^{-2}\bd_{1n}' (\bX_{1n}' (\hbbetaWR_{1n}-\hbbetaRE_{1n})(\hbbetaWR_{1n}-\bbeta^*_{1})'\bd_{1n}\} \\
 &=I_1+I_2+I_3.
 \end{array}
$$
From \eqref{aqr-ue}, we know $I_1=\lim_{n\to\infty} E\{n^{1/2}s_{1n}^{-1}\bd_{1n}'(\hbbetaWR_{1n}-\bbeta^*_{1})\}^2=1$.
From \eqref{beta1-new-residual},
$$
\begin{array}{ll}
I_2&=\lim_{n\to\infty} s_{1n}^{-2} E\{n^{1/2}\bd_{1n}'(\hbbetaWR_{1n}-\hbbetaRE_{1n})\}^2\\
&=\lim_{n\to\infty} s_{1n}^{-2} E\{n^{1/2}\bd_{1n}'\bSigma_{n11}^{-1}\bSigma_{n12}\hbbetaWR_{2n}\}^2\\
&=\lim_{n\to\infty}(s_{2n}^2/s_{1n}^{2}) E\{n^{1/2}s_{2n}^{-1}\bd_{2n}'\hbbetaWR_{2n}\}^2,
\end{array}
$$
where $\bd_{2n}=\bSigma_{n21}\bSigma_{n11}^{-1}\bd_{1n}$ and $s_{2n}^2=\bd_{2n}'\bSigma_{n22.1}^{-1}\bd_{2n}$.
From Ouellette (1981) Equation (1.12), we obtain
\bel{eq-lem-4}
\bSigma_{11}^{-1}\bSigma_{12}\bSigma_{22.1}^{-1}\bSigma_{21}\bSigma_{11}^{-1} =\bSigma_{11.2}^{-1} -\bSigma_{11}^{-1}.
\eel
Therefore,
$$s_{2n}^2=\sigma^2\bd_{1n}'\bSigma_{11}^{-1}\bSigma_{12}\bSigma_{22.1}^{-1}\bSigma_{21}\bSigma_{11}^{-1}\bd_{1n}
=\sigma^2\bd_{1n}'\bSigma_{11.2}^{-1}\bd_{1n} -\sigma^2\bd_{1n}'\bSigma_{11}^{-1}\bd_{1n}.$$
Since $s_{2n}^2/s_{1n}^2\to 1-c$,
$$I_2=(1-c) \lim_{n\to\infty} E[\chi^2_1(\Delta_{\bd_{1n}})]=(1-c)(1+\Delta_{\bd_{1n}}),$$
where $\chi_\nu^2(t)$ is a $\chi^2$ distribution with degrees of freedom $\nu$
and noncentral parameter $t$.
Here $\Delta_{\bd_{1n}}$ is given in \eqref{eq-Delta-d1n}.
From the Cauchy-Schwarz inequality,
$$\Delta_{\bd_{1n}}=s_{2n}^{-2}(\bd_{2n}'\bdelta)^2\le \bdelta'\bSigma_{n22.1}\bdelta.$$
Furthermore,
$$
\begin{array}{ll}
I_3=&-2 \lim_{n\to\infty} E \{nS_{1n}^{-2}\bd_{1n}' (\bX_{1n}' (\hbbetaWR_{1n}-\hbbetaRE_{1n})(\hbbetaWR_{1n}-\bbeta^*_{1})'\bd_{1n}\} \\
%&=2 \lim_{n\to\infty} E[\bd_{1n}'(\bI_{p_{1n}}~\bzero_{p_{1n}\times p_{2n}})
%nS_{1n}^{-2} (\hbbetaWR_{(-3)}-\hbbetaRE_{(-3)})(\hbbetaWR_{(-3)}-\hbbetaRE_{(-3)})'
%(\bzero_{p_{1n}\times p_{2n}}'~\bI_{p_{2n}})'
%\bSigma_{n21}\bSigma_{n11}^{-1}]\\
&=2 \lim_{n\to\infty} [\bd_{1n}'(\bI_{p_{1n}}~\bzero_{p_{1n}\times p_{2n}})
\bSigma_{n}^{-1}
(\bzero_{p_{1n}\times p_{2n}}'~\bI_{p_{2n}})'
\bSigma_{n21}\bSigma_{n11}^{-1}]\\
&=-2 \lim_{n\to\infty}(s_{2n}/s_{1n})^{2}=-2(1-c)
\end{array}
$$
Thus, $ R(\bd_{1n}'\hbbetaRE_{1n})=I_1+I_2+I_3=1-(1-\Delta_{\bd_{1n}})(1-c)$.
Thus \eqref{aqr-re} holds.

We now investigate \eqref{aqr-se}. First from the definition,
$$
\begin{array}{ll}
  R(\bd_{1n}'\hbbetaSE_{1n})&=\lim_{n\to\infty} E[n^{1/2}s_{1n}^{-1}\bd_{1n}'(\hbbetaSE_{1n}-\bbeta^*_{1}) ]^2\\
 &=  \lim_{n\to\infty}s_{1n}^{-2} E\{n^{1/2}\bd_{1n}'[(\hbbetaWR_{1n}-\bbeta^*_{1})-(\hbbetaWR_{1n}-\hbbetaRE_{1n})((p_{2n}-2)/T_n)]\}^2\\
 &=\lim_{n\to\infty} E\{n^{1/2}s_{1n}^{-2}\bd_{1n}'(\hbbetaWR_{1n}-\bbeta^*_{1})\}^2 -\\
   &\quad ( \lim_{n\to\infty} 2 E \{ ns_{1n}^{-2}((p_{2n}-2))/T_n)\bd_{1n}' (\hbbetaWR_{1n}-\hbbetaRE_{1n})(\hbbetaWR_{1n}-\bbeta^*_{1})'\bd_{1n}\}- \\
   &   \quad\quad  \lim_{n\to\infty}  E\{n^{1/2}s_{1n}^{-2}\bd_{1n}'(\hbbetaWR_{1n}-\hbbetaRE_{1n})((p_{2n}-2)/T_n) \}^2 )\\
 &=J_1-(J_2-J_3).
 \end{array}
$$
Again, $J_1=\lim_{n\to\infty} E\{n^{1/2}s_{1n}^{-2}\bd_{1n}'(\hbbetaWR_{1n}-\bbeta^*_{1})\}^2=1$. From \eqref{beta1-new-residual},
 $$\bd_{1n}'(\hbbetaWR_{1n}-\hbbetaRE_{1n})=-\bd_{1n}'\bSigma_{n11}^{-1}\bSigma_{n12}\hbbetaWR_{2n}=\bd_{2n}'\hbbetaWR_{2n}.$$
 Then we have
 $$
 \begin{array}{ll}
 J_2-J_3&= \lim_{n\to\infty} 2 E \{ n s_{1n}^{-2}((p_{2n}-2)/T_n)\bd_{1n}' (\hbbetaWR_{1n}-\hbbetaRE_{1n})(\hbbetaWR_{1n}-\bbeta^*_{1})'\bd_{1n}\} - \\
        &\quad              \lim_{n\to\infty}  E\{n^{1/2}s_{1n}^{-2}\bd_{1n}'(\hbbetaWR_{1n}-\hbbetaRE_{1n})((p_{2n}-2)/T_n) \}^2 \\
 &=-\lim_{n\to\infty}
             2 s_{1n}^{-2} E \{ ((p_{2n}-2)/T_n) \sqrt{n} \bd_{2n}'\hbbetaWR_{2n} \sqrt{n}(\hbbetaWR_{1n}-\bbeta^*_{1})'\bd_{1n}\}- \\
    &\quad   \lim_{n\to\infty} s_{1n}^{-2}  E\{  [((p_{2n}-2)/T_n) \sqrt{n} \bd_{2n}'\hbbetaWR_{2n}]^2\}
 \end{array}
 $$
 From Theorem  1,
$\widehat s_2=p_{2n}+o_p(1)$ and $$T_n=
(\sqrt{n}\hbbetaWR_{n2})'(\bSigma_{n22.1})(\sqrt{n}\hbbetaWR_{n2})/\hsigma^2+o_p(1).$$
%\end{equation}
We now define
 $\ba'=\begin{pmatrix}
 \bd_{1n}' & \bzero_{1\times p_{2n}}
 \end{pmatrix}
 $,
  $\bb'=\begin{pmatrix}
  \bzero_{p_{1n}\times 1}&
  -\bd_{2n}
 \end{pmatrix}
 $,
 and
 $
 \eta(\bx)=((p_{2n}-2)/(\bx'\bW\bx))\bb'\bx,$
 where
  $\bW=\begin{pmatrix}
 \bzero & \bzero\\
 \bzero & \bSigma_{n22.1}
 \end{pmatrix}.$
  Then from the asymptotic normality,
  $$
 J_2-J_3=\lim_{n\to\infty} s_{1n}^{-2}  E[2\eta(\bz+\bzeta) \ba'\bz -(\eta(\bz+\bzeta))^2 ],
 $$
 where $\bzeta'=\begin{pmatrix}
  \bzero_{p_{1n}\times 1}&
  -\bbeta_{20}'
 \end{pmatrix}
 $ and  $\bz$  satisfy that $$ (\bd_n'\bSigma_n^{-1/2} \bd_n)^{-1} \bd_{n}'\bz \to N(0,1)$$ and
 $$\lim_{n\to \infty} (\bd_n'\bSigma_n^{-1} \bd_n)^{-1}(\bd_n' \bzeta)^2=\lim_{n\to \infty}\Delta_{d_{1n}}.$$
 From Stein's lemma, we have
 $$
 \begin{array}{ll}
 E(\eta(\bz+\bzeta)\ba'\bz)
&=\ba'\bSigma_n^{-1}(\partial \eta(\bz+\bzeta)/\partial \bz)\\
 &= \dfrac{(p_{2n}-2)\ba'\bSigma_n^{-1}\bb}{(\bz+\bzeta)'\bW(\bz+\bzeta)}
       -\dfrac{2(p_{2n}-2) \ba'\bSigma_n^{-1}  \bW(\bz+\bzeta) (\bz+\bzeta)' \bb}{((\bz+\bzeta)'\bW(\bz+\bzeta))^2}
 \end{array}
 $$
 So we have
 $$
  \begin{array}{ll}
J_2-J_3&=\lim_{n\to\infty} s_{1n}^{-2}E[ 2\eta(\bz+\bzeta) \ba'\bz -(\eta(\bz+\bzeta))^2]\\
      & = \lim_{n\to\infty}s_{1n}^{-2}E \left\{\left[ 2\dfrac{(p_{2n}-2) \ba'\bSigma_n^{-1}\bb}{(\bz+\bzeta)'\bW(\bz+\bzeta)}
                    -4\dfrac{(p_{2n}-2) \ba'\bSigma_n^{-1}  \bW(\bz+\bzeta) (\bz+\bzeta)' \bb}{((\bz+\bzeta)'\bW(\bz+\bzeta))^2}  \right] \right\}\\
 &\quad  -\lim_{n\to\infty}s_{1n}^{-2} E  \left\{ \dfrac{(p_{2n}-2)^2  \bb'(\bz+\bzeta) (\bz+\bzeta)' \bb}{((\bz+\bzeta)'\bW(\bz+\bzeta))^2} \right\}\\
 &= \lim_{n\to\infty} E\left\{ \dfrac{(p_{2n}-2)}{(\bz+\bzeta)'\bW(\bz+\bzeta)} f
         \right\},
 \end{array}
 $$
 where
 $$f=  \frac{2\ba'\bSigma_n^{-1}\bb}{s_{1n}^{2}} -\dfrac{4 (\bz+\bzeta)' \bW \bSigma_n^{-1} \ba \bb' (\bz+\bzeta) }{s_{1n}^{2}(\bz+\bzeta)'\bW(\bz+\bzeta)}
         -\dfrac{(p_{2n}-2) (\bz+\bzeta)'\bb\bb' (\bz+\bzeta)}{s_{1n}^{2}(\bz+\bzeta)' \bW  (\bz+\bzeta)}.
$$
Notice that
$\ba'\bSigma_n^{-1}\bb=\bd_{2n}'\bSigma_{n22.1}^{-1}\bd_{2n}=s_{2n}^2$ and
$
 \bW \bSigma_n^{-1} \ba \bb'= \bb\bb' .
$
Therefore, $$f=2\dfrac{s_{2n}^2 }{s_{1n}^{2}}-\dfrac{(p_{2n}+2) (\bz+\bzeta)' \bb\bb'(\bz+\bzeta)}{s_{1n}^{2}(\bz+\bzeta)' \bW  (\bz+\bzeta)}.$$
Thus,
%\bel{eq-J23}
$$
\begin{array}{ll}
J_2-J_3&=\lim_{n\to\infty}E\left\{\dfrac{(p_{2n}-2)}{(\bz+\bzeta)'\bW(\bz+\bzeta)}\left[ \dfrac{2s_{2n}^2}{s_{1n}^{2}}-
          \dfrac{(p_{2n}+2) (\bz+\bzeta)' \bd_{2n}\bd_{2n}'(\bz+\bzeta)}{s_{1n}^2 (\bz+\bzeta)'\bW(\bz+\bzeta)}\right]\right\}\\
      &=\lim_{n\to\infty}\dfrac{s_{2n}^2 }{s_{1n}^{2}}
      E\left\{\dfrac{(p_{2n}-2)}{(\bz_2+\bdelta)'\bSigma_{n22.1}(\bz_2+\bdelta)}
      \left[ 2- \dfrac{(p_{2n}+2) (\bz_2+\bdelta)' \bd_{2n}\bd_{2n}'(\bz_2+\bdelta)}{s_{2n}^2(\bz_2+\bdelta)'\bSigma_{n22.1}(\bz_2+\bdelta) }\right]\right\},
\end{array}
$$
%\eel
 where $\bz_2$ satisfies that $ s_{2n}^{-1} \bd_{2n}'\bz_2 \to N(0,1)$.
 Thus \eqref{aqr-se} holds. Similarly, we can obtain \eqref{aqr-pse}.
$\Box$

\vspace{10pt}
\noindent {\bf Proof of Corollary \ref{cor3}.}

We first verify (i). Define $\widetilde\bz_2=\sigma^{-2}\bSigma_{n22.1}^{1/2}(\bz_2+\bdelta)$ and
$\bB=(\sigma^{2}/s_{2n}^2)\bSigma_{n22.1}^{-1/2}\bd_{2n}\bd_{2n}'\bSigma_{n22.1}^{-1/2}$.
%Notice that  the trace of $\bB$ is 1. Then from  ,
From the Cram${\rm\acute{e}}$r-Wold device, we have
%\bel{eq-J23-2}
$$
\begin{array}{ll}
J_2-J_3
&=(1-c)\lim_{n\to \infty} \left\{E\left[\dfrac{2(p_{2n}-2)}{\widetilde\bz_2'\widetilde\bz_2}\right]
-E\left[\dfrac{(p_{2n}-2)(p_{2n}+2) \widetilde\bz_2'\bB \widetilde\bz_2}{(\widetilde\bz_2'\widetilde\bz_2)^2}\right]\right\}\\
&=(1-c)\lim_{n\to \infty} \left\{E\left[\dfrac{p_{2n}-2}{\chi_{p_{2n}}^2(\Delta_n)}\right]
-{\rm Tr}(\bB )E\left[\dfrac{(p_{2n}-2)(p_{2n}+2) }{\chi_{p_{2n}+2}^4(\Delta_n)}\right]\right\}+\\
 &\quad (1-c)\lim_{n\to \infty} \left\{E\left[\dfrac{p_{2n}-2}{\chi_{p_{2n}}^2(\Delta_n)}\right] -
  (\bdelta'\bB\bdelta)E\left[\dfrac{(p_{2n}-2)(p_{2n}+2)}{\chi_{p_{2n}+4}^2(\Delta_n)}\right]  \right\},
\end{array}
$$
%\eel
where $\Delta_n=\bdelta'\bSigma_{n22.1}\bdelta$ and
 ``Tr($\bB$)'' is the trace of matrix $\bB$. Here the second ``='' is from
Theorem 8 in Chapter 2 in \cite{saleh:2006}.
Notice that Tr$(\bB)=1$. %and  $\bdelta'\bB\bdelta=\Delta_{\bd_{1n}}$.
 Using the relationship between
the Chi-square distribution and Poisson distribution, %\eqref{eq-J23-2} becomes
 $$
\begin{array}{ll}
J_2-J_3
&=(1-c) \lim_{n\to \infty} \left\{
     E_\kappa\left[\dfrac{p_{2n}-2}{p_{2n}-2+2 \kappa} \left(1-\dfrac{p_{2n}+2}{p_{2n}+2 \kappa} \right)\right]
     \right\} +\\
&\quad  (1-c) \lim_{n\to \infty} \left\{ E_\kappa\left[
         \dfrac{p_{2n}-2}{p_{2n}-2+2 \kappa}
        \left( 1-\dfrac{\bdelta'\bB\bdelta(p_{2n}-2+2\kappa)(p_{2n}+2)}{(p_{2n}+2+2 \kappa)(p_{2n}+2 \kappa)}\right)
      \right] \right\},
\end{array}
$$
where $\kappa$ is a poisson distribution with mean $\Delta_n/2$ and $E_\kappa$ means the expectation is
taken for the poisson random variable $\kappa$. Since $P(\kappa\ge 1) \to 1$ when $p_{2n} \to \infty$.
With almost probability 1, we have
$$
0\le \dfrac{p_{2n}-2}{p_{2n}-2+2 \kappa} \left(1-\dfrac{p_{2n}+2}{p_{2n}+2 \kappa}  \right)\le 1.
$$
If $\|\bdelta\|^2\le 1$, then
$\bdelta'\bB\bdelta=(\bdelta'\bSigma_{n22.1}^{-1/2}\bd_{2n})^2 / (\bd_{2n}'\bSigma_{n22.1}^{-1}\bd_{2n})
\le \bdelta'\bdelta\le 1$.
Then, $E[g_1(\bz_2+\bdelta)]\ge 0$. Furthermore, when
$\bx'\bSigma_{n22.1}\bx \le p_{2n}-2$, we have
$$
2-s_{2n}^{-2}\bx\bd_{2n}\bd_{2n}'\bx'\ge 2-\dfrac{ (p_{2n}-2)\bx\bd_{2n}\bd_{2n}'\bx'}{s_{2n}^{2}\bx'\bSigma_{n22.1}\bx }
> 2-\dfrac{ (p_{2n}+2)\bx\bd_{2n}\bd_{2n}'\bx'}{s_{2n}^{2}\bx'\bSigma_{n22.1}\bx } .
$$
Therefore, $g_2(\bx)\ge g_1(\bx)$.
Thus (i) holds.

In fact, the inequalities in (i) also  hold even though
$\|\bdelta\|^2>1$.
For example, suppose $\Delta_n=\iota p_{2n}$ for some constant $\iota>0$.
Then $p_{2n}^{-1/2}(2\kappa-\Delta_n)\leadsto N(0, \iota^{-1})$.
Therefore, if $\|\bdelta\|^2\le 1+\iota$, with probability 1 we have
$$
1-\dfrac{\bdelta'\bB\bdelta(p_{2n}-2+2\kappa)(p_{2n}+2)}{(p_{2n}+2+2 \kappa)(p_{2n}+2 \kappa)} \to 1-\dfrac{\|\bdelta\|^2}{1+\iota}>0.
$$
Thus, (ii) holds.

We now verify (iii).
 If $\bdelta=\bzero$,  then $\Delta_{\bd_{1n}}=0$.
 Thus, ${\rm ADR}(\bd_{1n}'\hbbetaRE_{1n} )=c<{\rm ADR}(\bd_{1n}'\hbbetaWR_{1n} )$.
We now compare ${\rm ADR}(\bd_{1n}'\hbbetaSE_{1n} )$ with ${\rm ADR}(\bd_{1n}'\hbbetaRE_{1n} )$.
Denote $\bA=(p_{2n}+2)s_{2n}^2\bSigma_{n22.1}^{-1/2}\bd_{2n}\bd_{2n}'\bSigma_{n22.1}^{-1/2}$.
If $\bdelta=\bzero$, then we have
$$
(1-c)^{-1}g_1(\bz_2)=\lim_{n\to\infty}\frac{(p_{2n}-2)}{\bz_2'\bSigma_{n22.1}\bz_2}-
\lim_{n\to\infty}\frac{(p_{2n}-2)(\bSigma_{n22.1}^{1/2}\bz_2)'\bA (\bSigma_{n22.1}^{1/2}\bz_2)}
{(\bz_2'\bSigma_{n22.1}\bz_2)^2}.
$$
From Theorem 2.1.8 in \cite{saleh:2006} and moment of inverse chi-squares distribution,
we have
$$
\lim_{n\to\infty} E\left[\frac{(p_{2n}-2)}{\bz_2'\bSigma_{n22.1}\bz_2}\right]=2
$$
and
$$
\lim_{n\to\infty}E\left[\frac{(p_{2n}-2)(\bSigma_{n22.1}^{1/2}\bz_2)'\bA (\bSigma_{n22.1}^{1/2}\bz_2)}
{(\bz_2'\bSigma_{n22.1}\bz_2)^2}\right]=\lim_{n\to\infty} 1+2/p_{2n}.
$$
Thus, if  $p_{2n}=p_2$ is fixed, $E[g_1(\bz_2)]=(1-c)(1-2/p_2)<1-c.$
Therefore,
$${\rm ADR}(\bd_{1n}'\hbbetaSE_{1n} )>{\rm ADR}(\bd_{1n}'\hbbetaRE_{1n} ).$$
Similarly, we can verify
${\rm ADR}(\bd_{1n}'\hbbetaRE_{1n} )<{\rm ADR}(\bd_{1n}'\hbbetaPSE_{1n} ).$
When $p_{2n}\to \infty$, ${\rm ADR}(\bd_{1n}'\hbbetaPSE_{1n})\to 1$.
$\Box$
%\end{appendix}

%\begin{ack}{ACKNOWLEDGEMENTS}
%We want to thank  the Associate Editor and two referees for their insightful comments on
% the improvement of  this paper.
%\end{ack}

\bibliographystyle{agsm}
\bibliography{allrefers-Gao20Dec2015}
%\bibliography{allrefers-Gao1Nov2015}

\end{document}